\documentclass[%
 reprint,
superscriptaddress,
 amsmath,amssymb,
 aps,
 pre, 
]{revtex4-2}

\usepackage{lipsum}
\usepackage{graphicx}

\usepackage{orcidlink} 
\usepackage{amsmath}

\usepackage{balance}
\usepackage{bm}
\newcommand{\bOmega}{\bm{\Omega}}
\newcommand{\bx}{\bm{x}}
\newcommand{\bq}{\bm{q}}
\newcommand{\be}{\bm{e}}

\newcommand{\Tbar}{\overline{T}}
\newcommand{\xbar}{\overline{x}}

\usepackage{epsfig}

\begin{document}

\title{Mean first passage time of chiral active Brownian particles}

\author{Sarafa A. Iyaniwura\,\orcidlink{0000-0002-8854-2335}}
\email{iyaniwura@aims.ac.za}
\affiliation{%
Vaccine and Infectious Disease Division,\\Fred Hutchinson Cancer Center, Seattle, Washington 98109, USA
}%

\author{Mingfeng Qiu\,\orcidlink{0000-0002-8114-4510}}
\email{mingfeng.qiu@canterbury.ac.nz}
\affiliation{School of Mathematics and Statistics, University of Canterbury, Christchurch 8140, New Zealand}

\author{Zhiwei Peng\,\orcidlink{0000-0002-9486-2837}}%
 \email{zhiwei.peng@ualberta.ca}
\affiliation{%
Department of Chemical and Materials Engineering,\\University of Alberta, Edmonton, Alberta T6G 1H9, Canada
}%

\date{\today}

\begin{abstract}
Chiral active Brownian particles (CABPs) are self-propelled agents with intrinsic rotational dynamics, giving rise to circular trajectories commonly observed in biological and synthetic microswimmers. Understanding how CABPs explore confined environments and locate targets is crucial for characterizing transport, search efficiency, and reaction processes in physical and biological systems. 
We study the escape dynamics of CABPs from one- and two-dimensional confined domains. In one dimension, we consider intervals with either two absorbing boundaries or a reflecting boundary on one side and an absorbing boundary on the other, and derive closed-form asymptotic solutions in the high-chirality regime, revealing the quantitative scaling of the mean first passage time (MFPT) as a function of  particle rotation speed (chirality). In two dimensions, we analyze escape from a disk containing one absorbing arc or two symmetric absorbing arcs. By numerically solving the governing partial differential equations, we compute the MFPT for CABPs to escape the domains as a function of the particle’s initial orientation, self-propulsion speed, angular velocity, and domain geometry.
Our results show that, depending on the parameters and geometry, the MFPT can exhibit non-monotonic behavior as a function of chirality. There exists an optimal chirality at an intermediate value that minimizes the escape time. Our work offers a comprehensive characterization of CABP escape dynamics in canonical confinements and identifies chirality as a key control parameter for transport and search in confined physical and biological systems.
\end{abstract}

\maketitle

\section{Introduction}

Chiral active Brownian particles (CABPs) are a class of self-propelled particles characterized by curved trajectories resulting from an angular velocity. Unlike standard active Brownian particles (ABPs) \cite{romanczuk2012active,solon2015active,zottl2023modeling}, CABPs break both time-reversal and parity symmetry, which fundamentally affects their mechanical response and transport behavior \cite{liebchen2022chiral,mecke2024emergent,soni2019odd,metselaar2019topological,liebchen2017collective,hargus2021odd,poggioli2023odd, langford2025phase,hargus2025flux}. This chiral motion arises from internal asymmetries within the particles themselves or from external torques due to magnetic fields or hydrodynamic interactions. Such dynamics has been observed across a range of systems, including certain motile bacteria, synthetic microswimmers, and colloidal particles near surfaces~\cite{PhysRevE.78.020101,upadhyaya2024narrow,PhysRevResearch.5.023196}.

CABPs can be used as a simple model for understanding active biological systems where chirality is important. Many microorganisms exhibit either intrinsic or emergent chiral swimming trajectories due to asymmetric flagellar motion, cell shape, and/or swimmer-boundary hydrodynamic interactions. These chiral dynamics enable microorganisms to navigate complex environments, respond to chemical gradients, and interact with surfaces in specialized ways~\cite{lauga2009hydrodynamics,magariyama2005difference}. Chiral motility has also been linked to biological processes such as biofilm formation, tissue organization, and immune cell migration~\cite{chin2018epithelial, rusconi2014microfluidics}. Insights from dynamics of CABPs can thus deepen our understanding of how motile cells exploit chirality for ecological fitness and adaptation. Moreover, these principles may inform the design of synthetic active systems for targeted drug delivery, microfluidic transport, and environmental sensing~\cite{bechinger2016active,chan2024chiral}.

Understanding transport and first-encounter statistics in active systems is central to many problems, from target search, chemotaxis, and infection spread to designing micro-robotic search strategies. The mean first passage time (MFPT), which quantifies the expected time for a particle to reach a target or exit a domain for the first time, is a fundamental metric for these processes. For achiral active Brownian particles (e.g., ABPs) exhibiting translational persistence, MFPT has been extensively studied, revealing how directional persistence, stochastic fluctuations, and domain geometry influence search efficiency~\cite{angelani2014first, moen2022trapping, scacchi2018mean,  baouche2025optimal, baouche2026spatiotemporal}. However, chirality qualitatively modifies trajectories, so first-passage statistics in that context often deviate from achiral  models.

Despite growing interest, quantitative theory for the MFPT of CABPs remains scarce. Existing studies have typically focused on passive and non-chiral active particles~\cite{iyaniwura2024asymptotic, di2023active, iyaniwura2025mean, locatelli2015active, baouche2026spatiotemporal}, and have not examined the effects of  chirality and rotational velocity---factors that can qualitatively alter search dynamics and first-passage behavior. Another relevant study compares ABPs and run-and-tumble particles near flat boundaries~\cite{moen2022trapping}, providing useful insights into boundary interactions, but such analyses likewise exclude chirality and do not extend to first-passage problems in more complex geometries. Environmental complexity has also been shown to strongly influence first-passage dynamics, as demonstrated in systems with passive crowders~\cite{biswas2020first}, yet these investigations remain largely focused on ABPs. A notable exception is recent work on the narrow escape problem for CABPs~\cite{upadhyaya2024narrow}. Using Brownian dynamics (BD) simulations, \citet{upadhyaya2024narrow} showed that the MFPT of CABPs is a non-monotonic function of chirality and exhibits a minimum at a finite chirality. While this represents an important step toward understanding first-passage phenomena in chiral systems, it focuses on specific boundary conditions and narrow escape scenarios, leaving open fundamental questions about how chirality modulates MFPTs across diverse geometries and parameter regimes. In particular, it remains unclear whether the observed optimal escape is specific to narrow escape settings or reflects a more general behavior of CABPs.

In this paper, we employ a backward Fokker–Planck equation to investigate the optimal escape dynamics of CABPs across a range of one- and two-dimensional geometries, with the goal of elucidating systematically how domain geometry and boundary conditions influence the escape of these particles. We begin by analyzing CABPs in one-dimensional  (1D) domains, first considering a finite interval bounded by two absorbing ends, which provides fundamental insight into how chirality and persistence shape escape times in the simplest setting. We then examine a modified 1D geometry in which one boundary is absorbing and the other reflecting, allowing us to isolate the influence of asymmetric boundary conditions on the escape time. For either case, we provide a complete asymptotic theory to characterize CABP escape behavior in 1D at high chirality, which has not yet been available from existing theoretical or computational results. At intermediate chirality, we use numerical simulations to study the escape times. Building on these results, we extend our analysis to two dimensions (2D) by considering CABPs escaping from a circular domain (a disk region) containing one or two openings on its boundary. In this context, we investigate how swim speed, chirality, and opening size modulate escape times, and how chirality interacts with geometric constraints to enhance or hinder escape.

Here we briefly outline the theoretical framework before detailed analysis. Let $P(\bx, \bq, t)$ denote the probability density of finding a CABP at position $\bx$ with orientation $\bq$ at time $t$; it is governed by the Fokker--Planck equation:
\begin{equation}
    \label{eq:FP}
    \frac{\partial P}{\partial t} = \mathcal{L}[P], 
\end{equation}
where $\mathcal{L}[P] = - \nabla \cdot\left( U_s \bq P\right) + D_x \nabla^2P - \nabla_R \cdot\left(\bOmega P\right) +D_R \nabla_R^2 P$. Here $U_s$ is the swim speed, $\bq$ denotes the swimming direction ($\bq\cdot\bq=1$), $D_x$ is the translational diffusivity, $\bOmega$ is the angular velocity, $D_R$ is the rotational diffusivity, $\nabla = \frac{\partial }{\partial \bx}$ is the translational gradient operator,  and $\nabla_R = \bq \times \frac{\partial }{\partial \bq}$ is the rotational gradient operator. The inverse of $D_R$ defines the reorientation time, $\tau_R=1/D_R$. If $\bOmega = \bm{0}$, one recovers the dynamics of ABPs, which is achiral.

We define the MFPT, \( T(\bx, \bq) \), as the mean time it takes for CABPs initially located at position $\bx$ with orientation  \( \bq \) to escape a domain. This quantity is governed by a backward Fokker--Planck equation, which reads
\begin{equation}
\label{eq:MFPT-EQ}
    \mathcal{L}^\dag[T] = -1,  
\end{equation}
where $\mathcal{L}^\dag = U_s \bq \cdot\nabla +D_x \nabla^2 +\bOmega \cdot \nabla_R + D_R \nabla_R^2$ is the adjoint operator of $\mathcal{L}$~\cite{zwanzig2001nonequilibrium}. Alternatively, one can obtain Eq.~\eqref{eq:MFPT-EQ} using a random walk approach~\cite{iyaniwura2025mean,iyaniwura2026splitting}.

\section{CABPs in one dimension}

In this section, we investigate the MFPT for CABPs escaping from a 1D interval. This setup models a narrow slit geometry in which translational motion is confined to the horizontal direction, while the orientation vector rotates unconstrained in two dimensions. We consider two boundary configurations: one with both boundaries absorbing and another with a reflecting left boundary and an absorbing right boundary. In both cases, we examine how the initial position  and orientation, swim speed, and chirality affect the escape time of the particles from the domain.

\subsection{CABPs in a 1D interval with absorbing boundaries at both ends}
\label{subsec:absorbing-1d}

\begin{figure}
    \centering
    \includegraphics[width=2.8in]{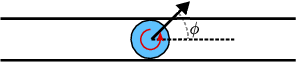}
    \caption{Schematic illustration of a chiral active Brownian particle (CABP) in a one-dimensional domain with absorbing boundaries at both ends. The black arrow indicates the self-propulsion direction $\bq$. The red curved arrow shows chirality. This setup models a narrow slit geometry in which the translational motion of the particle is restricted to the horizontal direction, while its orientation angle rotates continuously in 2D.}
    \label{fig:1d-absorb-schematic}
\end{figure}

Consider a CABP confined to a 1D interval $[-L,L]$, with absorbing boundaries at both ends, as illustrated in Fig.~\ref{fig:1d-absorb-schematic}. While its translational motion is confined to 1D, the CABP is allowed to freely rotate in 2D. We parametrize the orientation vector as $\bq = \cos \phi\; \be_x + \sin \phi \;\be_y$, where $\be_x$ and $\be_y$ are the unit basis vectors in the $x$ and $y$ directions, respectively. The angular velocity is defined as $\bOmega = \Omega \be_z$, where $\be_z = \be_x \times \be_y$ is the unit vector along the $z$ axis, perpendicular to the $x$-$y$ plane,  and $\Omega$ is the magnitude of the angular velocity. 

To make Eq.~\eqref{eq:MFPT-EQ} non-dimensional, we scale time by the swim timescale $\tau_s = L/U_s$ and length by $L$. The non-dimensional form of Eq.~\eqref{eq:MFPT-EQ} is given by 
\begin{equation}
\label{eq:T-non-dim-1D}
     \cos\phi\;  \frac{\partial \Tbar}{\partial \xbar} +  \frac{1}{Pe}\;\frac{\partial^2 \Tbar}{\partial \xbar^2} + \chi\;  \frac{\partial \Tbar}{\partial \phi} + \beta\; \frac{\partial ^2\Tbar}{\partial \phi^2}=-1, 
\end{equation}
where $\Tbar = T/\tau_s$, $\xbar = x/L$, $Pe = U_sL/D_x$,   $\chi = \Omega L/U_s$, and $\beta = L/(U_s\tau_R)$, with the MFPT $\Tbar \equiv \Tbar(\xbar, \phi)$ depending on the dimensionless initial position  $\xbar$  and orientation $\phi$ of the CABP.
Here, $Pe$ compares the diffusive timescale $L^2/D_x$ to the swim timescale $L/U_s$, $\chi$ compares the swim timescale with the rotational timescale $1/\Omega$, and $\beta$ represents the ratio of the swim timescale to the reorientation time $\tau_R$.  Equivalently, $\beta$ quantifies the ratio between the domain size $L$ and the persistence length $\ell=U_s\tau_R$.  Throughout this article, we refer to $\chi$ as the chirality  and only need to consider $\chi\geq 0$. The boundary conditions in $\xbar$ are given by $\Tbar (\pm 1, \phi)=0.$ Periodic boundary conditions are enforced for $\phi$: $\Tbar(\xbar, 0) = \Tbar(\xbar, 2\pi)$. 

\subsubsection{The high-chirality regime: two absorbing boundaries} \label{subsubsec:high-chirality-both-absorbing}

For high chirality, CABPs rotate rapidly and move in small circles because of the coupling to active swimming, which effectively reduces their swimming persistence. Therefore, in the limit $\chi \to \infty$, the MFPT is expected to approach that of passive Brownian particles (PBPs). To characterize the MFPT of CABPs in this high-$\chi$ regime, we employ a perturbation expansion given by
\begin{equation}\label{Eq:Perturb_Expand}
    \Tbar(\xbar, \phi) = \Tbar_0(\xbar, \phi) + \frac{1}{\chi} \; \Tbar_1(\xbar, \phi) + \cdots. 
\end{equation}
Substituting the expansion in \eqref{Eq:Perturb_Expand} into the PDE in \eqref{eq:T-non-dim-1D} and collecting terms in powers of $\chi$, at leading order $O(1)$, we have $\partial \Tbar_0/ \partial \phi=0$, which implies that $\Tbar_0$ is independent of the initial angular orientation of the particle and depends only on its initial position, i.e., $\Tbar_0 \equiv \Tbar_0(\xbar)$. At order $O(1/\chi)$, we obtain  
\begin{equation}\label{Eq:T0_1D_PDE}
     \cos \phi \frac{\partial \Tbar_0}{\partial \xbar} + \frac{1}{Pe}\; \frac{\partial ^2 \Tbar_0}{\partial \xbar^2}+ \frac{\partial \Tbar_1}{\partial \phi} +\beta \frac{\partial ^2 \Tbar_0}{\partial \phi^2} = -1.
\end{equation}
Integrating the PDE in Eq.~\eqref{Eq:T0_1D_PDE} over $\phi$ and applying the periodic boundary conditions for $\Tbar_{0}$ and $\Tbar_{1}$, we obtain 
\begin{equation}
\label{eq:passive}
    \frac{1}{Pe}\;\frac{\partial^2 \Tbar_0}{\partial \xbar^2} = -1,
\end{equation}
with corresponding boundary conditions given by $\Tbar_0(\xbar= \pm 1)=0$. We note that the appearance of $Pe$ in Eq.~\eqref{eq:passive} is due to  non-dimensionalization. In dimensional form, this PDE is equivalent to $D_x \partial^2 T_\mathrm{PBP} /(\partial x^2) = -1$, which  governs the MFPT of PBPs. 
This is consistent with the expected behavior of CABPs, whose dynamics recover those of PBPs in the limit $\chi \to \infty$.
The analytic solution to Eq.~\eqref{eq:passive} with the corresponding absorbing boundary conditions is
\begin{equation}
\label{eq:Tbar-passive}
    \Tbar_0(\xbar) = \frac{Pe}{2}\; (1-\xbar^2).
\end{equation}
In dimensional terms, $T_\mathrm{PBP}= \Tbar_0 \tau_s = \frac{1}{2}\frac{L^2}{D_x}\left(1 - (x/L)^2 \right)$.

Substituting $\Tbar_0$ into Eq.~\eqref{Eq:T0_1D_PDE} the $\partial^2 \Tbar_0/\partial \phi^2 = 0$ term varnishes. Solving the resulting equation for $\Tbar_1$, we obtain 
\begin{equation} 
\label{eq:Tbar1}
  \Tbar_1 = Pe \,\xbar\,\sin\phi,  
\end{equation}
where the integration constant vanishes by symmetry. Unlike the isotropic leading-order term $\Tbar_0$, this $O(1/\chi)$ bulk correction introduces an orientation-dependent escape time such that $\Tbar_1(\pm 1, \phi) = \pm\, Pe\,\sin\phi$ that violates the absorbing boundary conditions $\Tbar_1(\pm 1, \phi) = 0$. This mismatch implies the formation of boundary layers near the exits as $\chi \to \infty$. The boundary-layer thickness scales as  $O(1/\sqrt{\chi})$, which is set by the balance between translational diffusion and chiral rotation, $\frac{\partial ^2 \Tbar}{\partial \xbar^2}\sim  \chi \frac{\partial \Tbar}{\partial \phi}$. We therefore revise the outer bulk expansion given in Eq.~\eqref{Eq:Perturb_Expand} as 
\begin{equation}
\label{eq:bulk-expand-revised}
        \Tbar(\xbar, \phi) = \Tbar_0 + \frac{1}{\chi} \; \Tbar_1 + \frac{1}{\chi^{3/2}} \Tbar_{3/2}+ \frac{1}{\chi^2} \Tbar_2 +\cdots. 
\end{equation}
We note the introduction of an $O(1/\chi^{3/2})$ correction term  in the asymptotic expansion of the outer solution. This term is required to ensure consistent matching between the bulk and boundary layer solutions.
Substituting the expansion in Eq.~\eqref{eq:bulk-expand-revised} into Eq.~\eqref{eq:T-non-dim-1D}, we obtain the equation governing $\Tbar_{3/2}$: 
\begin{equation}
\label{eq:Tbar-half-orders}
   \frac{\partial \Tbar_{3/2}}{\partial \phi}=0. 
\end{equation}
The solutions to $\Tbar_0$ and $\Tbar_1$ are given in Eqs.~\eqref{eq:Tbar-passive} and \eqref{eq:Tbar1}, respectively. From Eq.~\eqref{eq:Tbar-half-orders}, we have $\Tbar_{3/2} \equiv \Tbar_{3/2}(\xbar)$. 
Collecting terms from Eq. \eqref{eq:T-non-dim-1D} at $O(1/\chi^{3/2})$ gives $\cos\phi\,\frac{\partial \Tbar_{3/2}}{\partial \xbar} + \frac{1}{Pe}\frac{\partial^2 \Tbar_{3/2}}{\partial \xbar^2} + \frac{\partial\Tbar_{5/2}}{\partial\phi} + \beta \frac{\partial^2 \Tbar_{3/2}}{\partial \phi^2} = 0.$ As before, integrating over $\phi$ and using the periodic boundary conditions for $\Tbar_{5/2}$, together with the fact that $\Tbar_{3/2}$ depends only on $\xbar$, yields $\Tbar_{3/2}'' = 0$.

To resolve the behavior near $\xbar=1$, we introduce a stretched boundary-layer coordinate $s=\sqrt{\chi} (1-\xbar)$.  Defining the boundary-layer solution as $\tilde{T}(s, \phi) = \Tbar(\xbar,\phi)$, we obtain 
\begin{equation}
\label{eq:absorbing-BL-eq}
\begin{split}
    -\frac{1}{\sqrt{\chi}}\,\cos \phi \,\frac{\partial \tilde{T}}{\partial s} + &\frac{1}{Pe}\frac{\partial^2 \tilde{T}}{\partial s^2} + \frac{\partial \tilde{T}}{\partial \phi} + \frac{\beta}{\chi} \frac{\partial^2 \tilde{T}}{\partial \phi^2} = - \frac{1}{\chi}, \\ 
    & \tilde{T}(0, \phi)=0. 
\end{split}
\end{equation}
In addition to the absorbing boundary, $\tilde{T}$ must match the bulk solution as $s \to + \infty$. Employing Van Dyke's matching rule~\cite{VanDyke1964}, we substitute Eqs.~\eqref{eq:Tbar-passive} and \eqref{eq:Tbar1} into the the bulk expansion in Eq.~\eqref{eq:bulk-expand-revised}, and  express the resulting solution in terms of the boundary-layer coordinate $s$, yielding
\begin{equation}\label{eq:absorbing-BLK-sol}
\begin{split}
        \Tbar =~&Pe\, s\,  \frac{1}{\sqrt{\chi}} +  Pe\left(\sin\phi-\frac{s^2}{2} \right) \frac{1}{\chi} \\ 
        &+\left(  - Pe\,s\,\sin\phi  + \Tbar_{3/2}(1)\right) \frac{1}{\chi^{3/2}}+ O(1/\chi^2),
\end{split}
\end{equation}
where $\Tbar_{3/2}(1) \equiv  \Tbar_{3/2}(\xbar=1) $. Observing the form of the bulk solution in Eq.~\eqref{eq:absorbing-BLK-sol}, we seek a boundary-layer solution of the form
\begin{equation}
\label{eq:absorbing1d-BL-expand}
    \tilde{T} = \frac{1}{\sqrt{\chi}} \tilde{T}_{1/2} + \frac{1}{\chi} \tilde{T}_1 + \frac{1}{\chi^{3/2}} \tilde{T}_{3/2} + \cdots, 
\end{equation}
where the matching conditions dictate that $\tilde{T}_{1/2} \sim Pe \, s$, $\tilde{T}_1 \sim  Pe\left(\sin\phi-\frac{s^2}{2} \right)$, and $\tilde{T}_{3/2} \sim - Pe\,s\,\sin\phi  + \Tbar_{3/2}(1)$ as $s \to \infty$.

Inserting the expansion in Eq.~\eqref{eq:absorbing1d-BL-expand} into Eq.~\eqref{eq:absorbing-BL-eq}, we obtain 
\begin{equation}
\begin{split}
 &\frac{1}{Pe}\frac{\partial^2 \tilde{T}_{1/2}}{\partial s^2} +\frac{\partial \tilde{T}_{1/2}}{\partial \phi}=0; \\ 
 &\tilde{T}_{1/2}(0,\phi)=0; \quad  \tilde{T}_{1/2}(s\to\infty,\phi) \sim Pe\,s,
\end{split}
\end{equation}
which has the solution $\tilde{T}_{1/2} = Pe\, s$. At $O(1/\chi)$, we have 
\begin{equation}
\label{eq:absorbing-T2-BL}
\begin{split}
 &-\cos\phi \frac{\partial\tilde{T}_{1/2}}{\partial s}+ \frac{1}{Pe}\frac{\partial^2 \tilde{T}_1}{\partial s^2} +\frac{\partial \tilde{T}_1}{\partial \phi}=-1;\\ 
 &\tilde{T}_1(0,\phi)=0; \quad \tilde{T}_1(s\to\infty,\phi)\sim  Pe\left(\sin\phi-\frac{s^2}{2} \right).
\end{split}
\end{equation}
We solve Eq.~\eqref{eq:absorbing-T2-BL} to obtain 
\begin{equation}\label{eq:T1_BL_sol}
\tilde{T}_1(s, \phi) = Pe \left( \sin\phi -\frac{s^2}{2}\right) - Pe\, e^{-\lambda s}\sin\left(\phi + \lambda s\right),
\end{equation}
where $\lambda = \sqrt{Pe/2}$. The first term  recovers the local expansion of the outer bulk solution near the exit, while the second term represents the boundary-layer correction. This correction takes the form of a damped oscillation that decays exponentially into the bulk with a characteristic dimensionless penetration depth of $\lambda^{-1}$.

At $O(1/\chi^{3/2})$, the governing equation is given by 
\begin{equation}
\label{eq:T-3halfs-eq}
    \begin{split}
        &-\cos\phi \frac{\partial\tilde{T}_{1}}{\partial s}+ \frac{1}{Pe}\frac{\partial^2 \tilde{T}_{3/2}}{\partial s^2} +\frac{\partial \tilde{T}_{3/2}}{\partial \phi} + \beta \frac{\partial^2 \tilde{T}_{1/2}}{\partial \phi^2}=0; \\ 
        &\tilde{T}_{3/2}(0,\phi)=0; \quad \tilde{T}_{3/2}(s\to\infty,\phi)\sim - Pe\,s\,\sin\phi  + \Tbar_{3/2}(1).
    \end{split}
\end{equation}
Substituting $\tilde{T}_{1/2} = Pe\, s$ and $\tilde{T}_{1}$ in Eq.~\eqref{eq:T1_BL_sol} into the PDE in Eq.~\eqref{eq:T-3halfs-eq}, and solving the resulting equation, we obtain 
\begin{equation}
\label{eq:T_3halfs}
\begin{split}
        \tilde{T}_{3/2}(s, \phi) = &\frac{Pe \lambda}{2} \Big[ H(\lambda s, 0) - 1 \Big]  - Pe\, s \sin\phi \\ 
        & + \frac{Pe \lambda}{2} \Big[ H(s\,\sqrt{Pe}, 2\phi) - H(\lambda s, 2\phi) \Big], 
\end{split}
\end{equation}
where we have introduced a damped function $H(\eta, \theta)$  that governs the boundary-layer decay:
\begin{equation}\label{eq:dampfunc_H}
    H(\eta, \theta) = e^{-\eta} \big[\cos(\eta + \theta) + \sin(\eta + \theta)\big].
\end{equation}
Noting that $H(\eta, \theta) \to 0$ as $\eta \to \infty$  and comparing Eq.~\eqref{eq:T-3halfs-eq} with Eq.~\eqref{eq:T_3halfs}, $\Tbar_{3/2}(1)$  is uniquely determined as $-Pe \lambda/2$. It is now clear that the presence of the  nonzero constant $-Pe \lambda/2$ in the limit $s\to \infty$ necessitates an $O(\chi^{-3/2})$ correction in the bulk region to satisfy the matching condition.

Owing to symmetry, the boundary-layer solutions near the left exit can be readily derived. Near $\xbar=-1$, we define $z =\sqrt{\chi}(1+\xbar)$ and obtain 
\begin{equation}
    \begin{split}
        &\tilde{T}_{1/2} =Pe\,z;\\ 
        &   \tilde{T}_1= -Pe \left( \sin\phi +\frac{z^2}{2}\right) + Pe\, e^{-\lambda z}\sin\left(\phi + \lambda z\right); \\ 
        & \tilde{T}_{3/2} = \frac{Pe \lambda}{2} \Big[ H(\lambda z, 0) - 1 \Big]  + Pe\, z \sin\phi \\ 
        & + \frac{Pe \lambda}{2} \Big[ H(z\,\sqrt{Pe}, 2\phi) - H(\lambda z, 2\phi) \Big]. 
    \end{split}
\end{equation}
From this, we have $\Tbar_{3/2}(-1) =-Pe\lambda/2$. Since $\Tbar_{3/2}(\pm 1) = -Pe\lambda/2$, the equation $\Tbar_{3/2}^{\prime\prime}=0$ admits a constant solution given by $\Tbar_{3/2} = -Pe \lambda /2$. 

\begin{figure}[!h]
    \centering
    \includegraphics[width=3in]{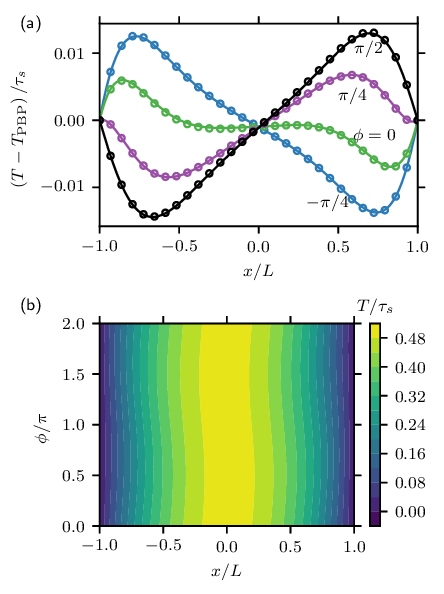}
    \caption{ Asymptotic solutions for the MFPT of CABPs in a 1D interval with absorbing boundaries.  (a) Comparison between the full numerical (FEM; solid lines) and asymptotic (markers) solutions for  $\left(T-T_\mathrm{PBP}\right)/\tau_s$. The plotted asymptotic solution [see Eq.~\eqref{eq:absorbing-composite}] is given by $\left(T-T_\mathrm{PBP}\right)/\tau_s = F_1/\chi +F_{3/2}/\chi^{3/2}$. (b) Contour plot showing the MFPT obtained from the three-term asymptotic solution in Eq.~\eqref{eq:absorbing-composite}. For both (a) and (b), $Pe=1$, $\chi=50$, and $\beta=0.1$. }
    \label{fig:1d-absorbing-theory}
\end{figure}

Using the bulk and boundary-layer solutions, we construct a composite asymptotic solution as 
\begin{equation}
\label{eq:absorbing-composite}
\begin{split}
      \Tbar_c = &\frac{Pe}{2}\left( 1-\xbar^2\right) + \frac{1}{\chi}F_1 +\frac{1}{\chi^{3/2}} F_{3/2} + O(1/\chi^2), 
\end{split}
\end{equation}
where 
\begin{equation}
\label{eq:F1}
    F_1=Pe\left[ \xbar\sin\phi -e^{-\lambda s} \sin \left(\phi +\lambda s\right)+ e^{-\lambda z} \sin \left(\phi+\lambda z \right)\right],
\end{equation}
\begin{equation}
\label{eq:F2}
\begin{split}
    F_{3/2}=  \frac{Pe \lambda}{2} \Bigl[& -1+H(\lambda s, 0) + H(s\,\sqrt{Pe}, 2\phi) - H(\lambda s, 2\phi) \\ 
    &+ H(\lambda z, 0) + H(z\,\sqrt{Pe}, 2\phi) - H(\lambda z, 2\phi) \Bigr].
\end{split}
\end{equation}
In Eqs.~\eqref{eq:F1} and \eqref{eq:F2},  $s = \sqrt{\chi}(1-\xbar)$, $z = \sqrt{\chi}(\xbar+1)$, $\lambda = \sqrt{Pe/2}$, and the function $H(\eta, \theta)$ is as defined in Eq.~\eqref{eq:dampfunc_H}. We note that the asymptotic solution in Eq.~\eqref{eq:absorbing-composite} is independent of $\beta$.
For particles initially located at the midpoint of the interval ($\xbar = 0$), the $Pe\,\xbar\,\sin\phi$ term in Eq.~\eqref{eq:F1} vanishes,  yielding
\begin{equation}
\label{eq:Tbar-midpoint-absorbing}
    \Tbar_c(0,\phi) = \frac{Pe}{2}\left(1 - \frac{\lambda}{\chi^{3/2}}\right) +\cdots,
\end{equation}
where exponentially small terms are neglected. Thus,  $\Tbar(0,\phi)$ is independent of $\phi$, at least up to $O(1/\chi^{3/2})$. In the bulk, for $\xbar=0$ or $\phi=0, \pi$, the first correction to the passive result is  at $O(1/\chi^{3/2})$, whereas for $\xbar\neq 0$ and $\phi \neq 0, \pi$, it is at $O(1/\chi)$.

To validate the asymptotic solution in Eq.~\eqref{eq:absorbing-composite}, we solve the full problem [Eq.~\eqref{eq:T-non-dim-1D}] numerically  using a finite element method (FEM) implemented in \texttt{FreeFem++}.\cite{hecht2012new} We compute the MFPT from FEM for $\chi=50$, $Pe=1$, and $\beta=0.1$.  In Fig. \ref{fig:1d-absorbing-theory}(a), we compare the correction terms beyond leading order, namely $(T-T_\mathrm{PBP})/\tau_s = F_1/\chi + F_{3/2}/\chi^{3/2}$ (solid lines), with those obtained from FEM (markers). For FEM, the corresponding result is defined as $(T_\mathrm{num} - T_\mathrm{PBP})/\tau_s$, where $T_\mathrm{num}$ denotes the full numerical solution of the MFPT. As shown in Fig. \ref{fig:1d-absorbing-theory}(a), the asymptotic solution agrees very well with the full numerical solution over the entire $x$-domain for different values of $\phi$.

In Fig. \ref{fig:1d-absorbing-theory}(a), we present the MFPT correction relative to PBPs for several initial orientations $\phi$ as a function of initial location $\xbar = x/L$. Owing to the symmetry of the domain and equations, it suffices to consider $0 < \xbar < 1$ and $-\pi/2 < \phi <\pi/2$. As $\phi$ increases from negative to positive values, CABPs transition continuously from facilitated to hindered transport regimes. For $\chi > 0$, CABPs rotate counterclockwise. Although the particle initially sits closer to the right exit ($\xbar >0$), for $\phi > 0$ the rotation steers it further upward, which may cause it to miss the right boundary. As a result, escape through the right exit is generally delayed. On the other hand, for $\phi < 0$, the particle’s orientation is further steered toward the right exit as the particle approaches the exit, thereby facilitating escape. We note that there is a delicate balance between chirality, swim-induced escape, and Brownian diffusion. Taking $\phi = 0$ as an example, propulsion-accelerated escape is only effective when the particle is sufficiently close to the exit, otherwise the chirality-induced rotation may oversteer the particle's orientation before it reaches the boundary. In that case, the particle has not yet escaped while its orientation has already reversed ($\phi > \pi/2$). If the particle starts exactly at the exit, MFPT is zero. As a result, for a fixed $\phi$, the quantity $(T - T_\mathrm{PBP})/\tau_s$ exhibits a non-monotonic dependence on $\xbar$ [see Fig.~\ref{fig:1d-absorbing-theory}(a)].

At the midpoint of the interval, the asymptotic solution [at least up to $O(1/\chi^{3/2})$] in this high-chirality regime is independent of $\phi$; therefore, the curves for different $\phi$ intersect at $\xbar =0$. From Eq.~\eqref{eq:Tbar-midpoint-absorbing}, it follows that $(T-T_\mathrm{PBP})/\tau_s \sim -Pe\lambda/(2\chi^{3/2})\approx -0.0047$ for $Pe=1$ and $\chi=50$ [Fig.~\ref{fig:1d-absorbing-theory}(a)].

Our asymptotic theory and Fig.~\ref{fig:1d-absorbing-theory}(a) highlight the fine structure of the dependence of the MFPT of CABPs on the initial orientation angle of the particles in the high chirality regime, although this dependence is generally weak. Notice the small magnitude of the correction to the MFPT evidenced in Fig.~\ref{fig:1d-absorbing-theory}(a). Indeed the MFPT of CABPs is, to leading order, identical to that of PBPs [see first term in Eq.~\eqref{eq:absorbing-composite}]. This is demonstrated in Fig. \ref{fig:1d-absorbing-theory}(b) where we show a contour plot of the MFPT for CABPs obtained from the three-term asymptotic solution in Eq.~\eqref{eq:absorbing-composite} as a function of the initial position and orientation of the particles, for $\chi=50$ and $Pe=1$. Here the MFPT is normalized by the constant swimming time scale $\tau_s$ to provide a global view of the actual magnitude of escape time. As expected, overall escape is slower when a particle starts closer to the midpoint.

\subsubsection{The finite-chirality regime: two absorbing boundaries}

\begin{figure*}[tb]
    \centering
    \includegraphics[width=6.5in]{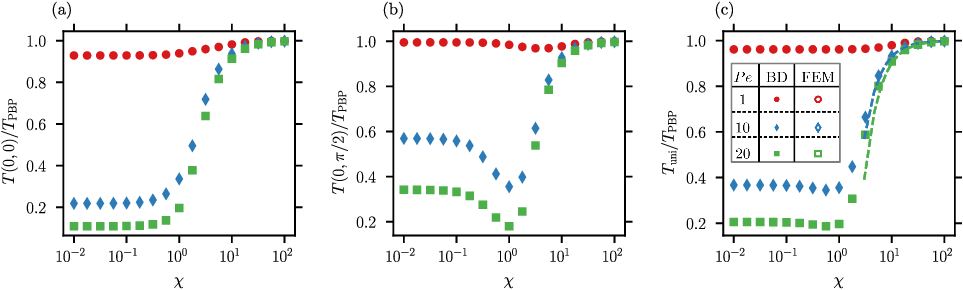}
    \caption{Scaled MFPT as a function of chirality ($\chi$) for CABPs in a 1D interval with absorbing boundaries, computed using BD (see Appendix~\ref{app:BD}) and FEM. We plot $T(\xbar, \phi)$ normalized by the passive MFPT $T_\mathrm{PBP}$, i.e., $T/T_\mathrm{PBP}$, for particles starting at $x = 0$ under three initial-orientation conditions: (a) $\phi = 0$ (right-pointing), (b) $\phi = \pi/2$ (orientation perpendicular to the exits), and (c) $\phi$ uniformly random, and for three swim speed: $Pe = 1$ (red circles), $Pe = 10$ (blue diamonds), and $Pe = 20$ (green squares).
 In each case, $T_{\mathrm{PBP}}$ is computed from Eq.~\eqref{eq:Tbar-passive}. For all results, $\beta=0.1$. The dashed lines indicate the high-chirality solution given in Eq.~\eqref{eq:absorbing-large-chi-scaled}. The FEM and BD solutions have excellent agreement and cannot be distinguished visually.}
    \label{fig:mfpt-1d-absorbing}
\end{figure*}

\begin{figure*}
    \centering
    \includegraphics[width=6.5in]{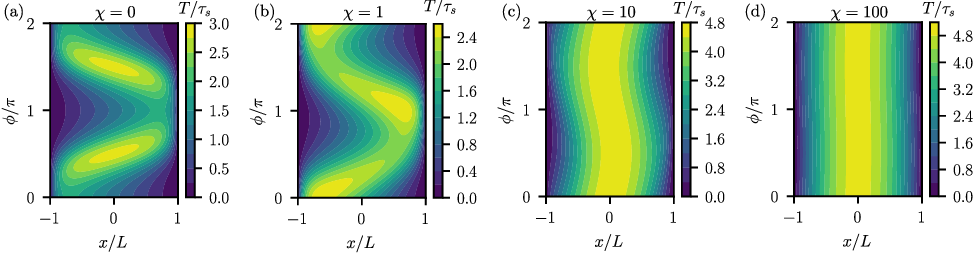}
    \caption{Contour plots showing the MFPT of CABPs in a 1D interval with absorbing boundaries as a function of the initial position ($x/L$) and orientation ($\phi/\pi$) for (a) $\chi = 0$, (b) $\chi = 1$, (c) $\chi = 10$, and (d) $\chi = 100$. In all cases, $Pe = 10$ and $\beta = 0.1$.}
    \label{fig:absorbing-contour}
\end{figure*}

We next study the MFPT for CABPs on the 1D interval $[-1,1]$ with absorbing boundaries in the finite-chirality regime, for which no closed-form analytical solution is available. This regime also includes the limit $\chi \to 0$, which recovers the ABP case. For arbitrary $\chi$, we therefore solve Eq.~\eqref{eq:T-non-dim-1D} numerically using a finite element method (FEM) implemented in \texttt{FreeFem++}~\cite{hecht2012new}. Results are presented in Figs.~\ref{fig:mfpt-1d-absorbing} and \ref{fig:absorbing-contour}. To further validate our results, for these cases, we also carry out Brownian Dynamics (BD; see Appendix~\ref{app:BD}) simulations  and compare the results with the FEM solution. Excellent agreement has been achieved, and thus we omit BD simulations for the scenarios studied in the rest of the paper.

In Fig.~\ref{fig:mfpt-1d-absorbing}, we show the MFPT for CABPs normalized by that of PBPs, i.e., $T/T_\mathrm{PBP}$, for particles starting at $\xbar = 0$ as a function of $\chi$. Results are shown for varying P\'eclet numbers ($Pe = 1$, $Pe = 10$, and $Pe = 20$) and initial orientations: right-pointing [$\phi = 0$, Fig.~\ref{fig:mfpt-1d-absorbing}(a)], perpendicular to the exits [$\phi = \pi/2$, Fig.~\ref{fig:mfpt-1d-absorbing}(b)], and uniformly random $\phi$ [Fig.~\ref{fig:mfpt-1d-absorbing}(c)]. For uniformly random initial orientations, we define
\begin{equation}
    T_\mathrm{uni} =  \frac{1}{2\pi} \int_0^{2\pi} T(0, \phi)\text{d} \phi. 
\end{equation}
For ABPs ($\chi=0$) and CABPs with weak chirality, the MFPT decreases with increasing $Pe$ since particles can exit from both ends of the interval. Increasing $Pe$ corresponds to a lower translational diffusivity, thereby  enhancing escape driven by self-propulsion. In contrast, when $\chi$ becomes large, the MFPT converges to that of PBPs at leading-order for all values of $Pe$, consistent with our asymptotic analysis in the high-chirality regime (Section~\ref{subsubsec:high-chirality-both-absorbing}). Using Eq.~\eqref{eq:Tbar-midpoint-absorbing}, we obtain the analytical solution
\begin{equation}
\label{eq:absorbing-large-chi-scaled}
    \frac{T}{T_\mathrm{PBP}}\Big\rvert_{\xbar=0} = 1 - \frac{\lambda}{\chi^{3/2}} +\cdots, \quad \mathrm{as}~\chi \to \infty,
\end{equation}
where $\lambda = \sqrt{Pe/2}$. For $Pe=10$ and $Pe=20$, this large-$\chi$ limit is indicated by the dashed lines  in Fig. \ref{fig:mfpt-1d-absorbing}(c).

For a fixed $Pe$, the MFPT increases monotonically with $\chi$ for right-pointing particles [see Fig. \ref{fig:mfpt-1d-absorbing}(a)]. Physically, increasing $\chi$ enhances the rate of rotation, thereby reducing the effective swimming persistence. As a result, particle escape mediated by self-propulsion is increasingly suppressed. We note that, by symmetry, left-pointing particles exhibit the same MFPT behavior as right-pointing particles. 

In contrast, for particles initially oriented perpendicular to the exits [$\phi=\pi/2$; see Fig. \ref{fig:mfpt-1d-absorbing}(b)], there exists an intermediate value of $\chi$ that minimizes the MFPT. In this scenario, the MFPT initially decreases as $\chi$ increases, reaches a minimum, and then increases for larger $\chi$ values. For $\chi=0$, ABPs oriented toward the wall typically spend a time of order $\tau_R$ reorienting before they are able to swim toward the absorbing boundary. When $\chi$ is small, particles can  acquire a non-zero horizontal velocity, allowing them to escape more quickly compared to ABPs. In the limit $\chi \to \infty$,  self-propulsion no longer contributes to directed motion, and escape occurs purely via Brownian motion. An optimum is obtained when the swim timescale is comparable to the timescale of rotation due to chirality, $L/U_s \approx 1/\Omega$, which gives $\chi \approx 1$.

When the initial orientation of the particles is uniformly random, the MFPT profile represents
an average of the MFPTs for right-pointing particles and for particles oriented perpendicular to 
the boundaries. Upon averaging over the orientations [see Fig. \ref{fig:mfpt-1d-absorbing}(c)], the non-monotonicity in $\chi$ becomes much less pronounced compared to Fig. \ref{fig:mfpt-1d-absorbing}(b) for top-pointing particles.

While uniformly random initial orientations may better represent realistic scenarios, it is instructive to examine how the MFPT depends more generally on the initial orientation. Fig.~\ref{fig:absorbing-contour} shows contour plots of the MFPT for CABPs in a 1D interval with absorbing boundaries as a function of the initial position and orientation of the particles, for different chirality. In all cases, the swim speed is fixed at $Pe = 10$, and $\beta = 0.1$. 
When $\chi = 0$ (achiral), the MFPT is maximized when particles start at the center of the domain ($\xbar = 0$) and are oriented perpendicularly to the boundaries, with the MFPT profile symmetric about $\phi = \pi$. As the starting position moves toward the right boundary ($\xbar = 1$), the MFPT increases for orientations pointing toward the left boundary ($\phi \approx \pi$) [see Fig.~\ref{fig:absorbing-contour}(a)]. Conversely, for particles starting near the left boundary, the MFPT increases for orientations pointing toward the right boundary, with the symmetry preserved about $\phi = \pi$. As expected for $\chi = 0$, particles starting very close to a boundary exhibit the maximum MFPT when initially oriented directly toward the opposite boundary.

A similar trend is observed for $\chi = 1$ [Fig.~\ref{fig:absorbing-contour}(b)], which yields the lowest maximum MFPT among the values of $\chi$ considered  in Fig.~\ref{fig:absorbing-contour}. As $\chi$ increases further, the MFPT maximum remains near the center of the domain but becomes less sensitive to the initial orientation [Figs.~\ref{fig:absorbing-contour}(c)--(d)]. For large $\chi$, the MFPT converges to that of PBPs, as shown for $\chi = 100$ in Fig.~\ref{fig:absorbing-contour}(d), where it is maximized at $\xbar = 0$ and decreases monotonically and symmetrically as the starting position approaches either boundary, independent of orientation.

\subsection{CABPs in a 1D interval with a reflecting left boundary}\label{sec:MFPT_1D}

\begin{figure}
    \centering
    \includegraphics[width=2.8in]{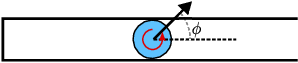}
    \caption{Schematic illustration of a chiral active Brownian particle (CABP) in a one-dimensional domain with a reflecting left boundary. The blue circle represents the CABP. The black arrow indicates the orientation vector. The red curved arrow shows chirality.}
    \label{fig:1d-mixed-schematic}
\end{figure}

Extending the previous case of absorbing boundaries at both ends, we now consider a CABP confined to the 1D interval $[-L, L]$, with a reflecting boundary at the left end and an absorbing boundary at the right end, as illustrated in Fig.~\ref{fig:1d-mixed-schematic}. In this configuration, the reflecting boundary prevents escape to the left, so the particle can only exit the domain through the right absorbing boundary.  Our results highlight the effect of asymmetry in the confinement geometry on CABP dynamics.

The MFPT for the CABP to escape the domain in this configuration is parameterized analogously to that of the symmetric interval with absorbing boundaries at both ends. Consequently, the governing non-dimensional PDE for the MFPT is identical to Eq.~\eqref{eq:T-non-dim-1D}, but with boundary conditions adapted to the mixed (reflecting--absorbing) domain geometry:  
\begin{equation}\label{Eq:BC_reflect_Absorb}
    \frac{\partial \Tbar}{\partial \xbar}(-1, \phi)=0, 
    \quad \text{and} \quad \Tbar(1, \phi)=0,
\end{equation}
where the reflecting condition at $\xbar = -1$ enforces a zero flux of probability at the left boundary, and the absorbing condition at $\xbar = 1$ indicates certain escape upon reaching the right boundary. We also impose a periodic boundary condition for $\phi$: $\Tbar(\xbar, 0) = \Tbar(\xbar, 2\pi)$.

\subsubsection{The high-chirality regime: reflecting-absorbing boundaries\label{subsubsec:reflect-absorb-theory}}

Similar to the case with absorbing boundaries at both ends, we employ matched asymptotic expansions to derive analytical solutions in the high-chirality limit. In the bulk, we use the regular perturbation expansion introduced in Eq.~\eqref{Eq:Perturb_Expand}. This leads to the same leading-order problem for $\Tbar_0$ as in Eq.~\eqref{eq:passive}, but now solved with the boundary conditions in Eq.~\eqref{Eq:BC_reflect_Absorb}. Solving this boundary-value problem (BVP) yields an explicit expression for the leading-order MFPT in the limit $\chi \to \infty$:  
\begin{equation}\label{eq:Tbar-passive_Ref_Abs_BC}
    \Tbar_0(\xbar) = \frac{Pe}{2}\,(\xbar+3)(1-\xbar),
\end{equation}
which is identical to the MFPT of PBPs in this interval.  Substituting $\Tbar_0$  into Eq.~\eqref{Eq:T0_1D_PDE} and solving, we obtain 
\begin{equation}
\label{eq:T1-bulk-leftreflect}
    \Tbar_1 = Pe \left(1 + \xbar \right)\sin\phi + a(\xbar),
\end{equation}
where $a(\xbar)$ is an unknown function of $\xbar$.

Near the right absorbing boundary, we again introduce the stretched boundary-layer coordinate $s = \sqrt{\chi}(1-\xbar)$. We express the bulk expansion $\Tbar_0 + \Tbar_1/\chi +\cdots$ in terms of the boundary-layer coordinate $s$ and expand it in a series as $\chi \to \infty$, yielding
\begin{equation}
\label{eq:matching-RBL-reflec}
        \Tbar =  2 Pe\, s\, \frac{1}{\sqrt{\chi}} + \left[  a(1)-\frac{Pe \, s^2}{2}+2 Pe\, \sin \phi   \right]\frac{1}{\chi} + O\left(1/\chi^{3/2}\right),
\end{equation}
where $a(1) \equiv a(\xbar = 1)$.
The boundary-layer equation and asymptotic expansion are identical to those in the case where both boundaries are absorbing; they are given by Eqs.~\eqref{eq:absorbing-BL-eq} and \eqref{eq:absorbing1d-BL-expand}, respectively. The distinction lies solely in the matching conditions: $\tilde{T}_{1/2} \sim 2 Pe\, s$ and $\tilde{T}_1 \sim a(1) - Pe\, s^2/2+2\,Pe\,\sin\phi$ as $s \to \infty$. Solving the boundary-layer equations subject to these matching conditions yields
\begin{equation}
\label{eq:RBL-left-reflect-sols}    
\begin{split}
    &\tilde{T}_{1/2} = 2Pe\,s;\\ 
    &\tilde{T}_1 = - \frac{Pe\,s^2}{2}+2Pe\left( \sin\phi - e^{-\lambda s} \sin(\lambda s +\phi)\right).
\end{split}
\end{equation}
A necessary condition for the above solution is that $a(1)=0$. 

\begin{figure}[!h]
    \centering
    \includegraphics[width=3in]{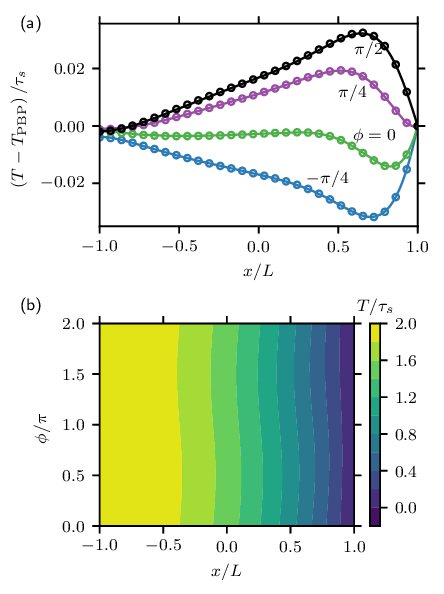}
    \caption{ Asymptotic solutions for the MFPT of CABPs in a 1D interval with reflecting left and absorbing right boundaries, .  (a) Comparison between the full numerical (FEM; solid lines) and asymptotic (markers) solutions for  $\left(T-T_\mathrm{PBP}\right)/\tau_s$. The plotted asymptotic solution [see Eq.~\eqref{eq:reflect-composite}] is given by $\left(T-T_\mathrm{PBP}\right)/\tau_s = \left( G_1 + \Tbar_1\right)/\chi +\left( G_{3/2}+\Tbar_{3/2}\right)/\chi^{3/2}$. (b) Contour plot showing the MFPT obtained from the three-term asymptotic solution in Eq.~\eqref{eq:reflect-composite}. For both (a) and (b), $Pe=1$, $\chi=50$, and $\beta=0.1$. The asymptotic and full solutions in panel (a) have excellent agreement. }
    \label{fig:1d-reflect-theory}
\end{figure}

Near the left reflecting boundary, we introduce the stretched boundary-layer coordinate $z = \sqrt{\chi}(1+\xbar)$. Similar to Eq.~\eqref{eq:matching-RBL-reflec}, we express the outer solution in terms of  $z$ to obtain
\begin{equation}
\label{eq:reflect-left-matching}
    \Tbar = 2 Pe + \left[ a(-1) - \frac{1}{2}Pe\;  z^2\right] \frac{1}{\chi} + O\left( 1/\chi^{3/2}\right),
\end{equation}
 This suggests a boundary-layer expansion of the form
\begin{equation}
\label{eq:reflecting-BL-expand}
    \tilde{T} =  \tilde{T}_{0} + \frac{1}{\chi} \tilde{T}_1 +  \frac{1}{\chi^{3/2}} \tilde{T}_{3/2} + \cdots, 
\end{equation}
with matching conditions $\tilde{T}_{0} \sim 2Pe$ and $\tilde{T}_1 \sim a(-1) - Pe\, z^2/2$ 
as $z \to \infty$. Substituting Eq.~\eqref{eq:reflecting-BL-expand} into the boundary-layer 
PDE~\eqref{eq:absorbing-BL-eq} and solving each order in turn, subject to these matching 
conditions, yields
\begin{equation}
\label{eq:T01-lbl-reflect}
    \tilde{T}_0 = 2\, Pe; \quad \tilde{T}_1= -\frac{1}{2} Pe\, z^2,
\end{equation}
which additionally requires $a(-1) = 0$. Solving the boundary value problem (BVP) $a^{\prime\prime} = 0$ subject to $a(\pm 1) = 0$ 
then gives $a(\bar{x}) = 0$.

At $O(1/\chi^{3/2})$, the governing equation for the right boundary layer remains the same as Eq.~\eqref{eq:T-3halfs-eq}. 
Substituting the solutions in Eq.~\eqref{eq:RBL-left-reflect-sols} into Eq.~\eqref{eq:T-3halfs-eq} and solving the resulting equation, we obtain the solution for $\tilde{T}_{3/2}$ at the right boundary layer, given by 
\begin{equation}
\begin{split}
        \tilde{T}_{3/2}(s, \phi) =&~ Pe\lambda\Big[ H(\lambda s, 0) - 1 \Big] - Pe\,s\,\sin\phi \\ 
        & + Pe\lambda \Big[ H( s \sqrt{Pe}, 2\phi) - H(\lambda s, 2\phi) \Big],
\end{split}
\end{equation}
where the function $H(\eta, \theta)$ is as defined in \eqref{eq:dampfunc_H}. As $s \to \infty$, in outer variables, we have $\Tbar_{3/2}(1) = -Pe \lambda$.
Similarly, at the left boundary layer, the solution at $O(1/\chi^{3/2})$ is 
\begin{equation}
\label{eq:T32-reflect-lbl}
    \tilde{T}_{3/2}(z,\phi) = Pe \, z \sin \phi + \Tbar_{3/2}(-1) + \lambda  H(\lambda  z,\phi ).
\end{equation}

We note that the left solution given in Eq.~\eqref{eq:T32-reflect-lbl} still contains the undetermined constant $\Tbar_{3/2}(-1)$. 
To solve the BVP $\Tbar_{3/2}^{\prime\prime}=0$ [cf. Eq.~\eqref{eq:Tbar-half-orders}], we need an additional boundary condition. In the Appendix, we show that this condition is provided by the solvability condition for the left boundary layer at the next order. The solvability requires that $\Tbar_{3/2}^\prime(-1) = Pe \;\lambda /2$.  With this, one can readily solve the BVP to obtain 
\begin{equation}\label{eq:T32-reflect-sol}
    \Tbar_{3/2}(\xbar) = \frac{Pe \; \lambda}{2}\left( \xbar - 3\right).
\end{equation}
We then calculate the value of $\Tbar_{3/2}$ at the left boundary as $\Tbar_{3/2}(-1) = -2 Pe\;\lambda$, which uniquely determines the solution given in Eq.~\eqref{eq:T32-reflect-lbl}.

\begin{figure*}
    \centering
    \includegraphics[width=6.5in]{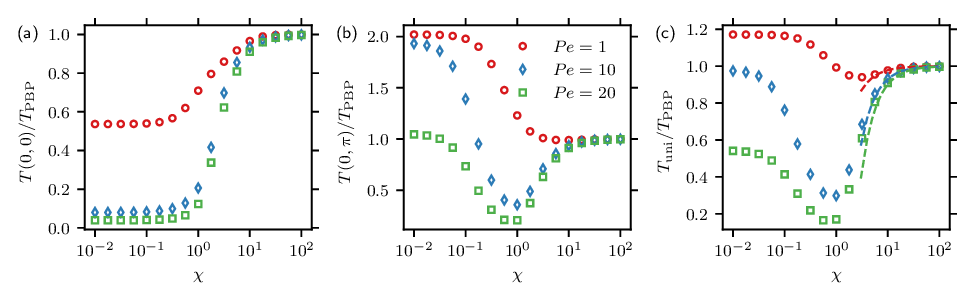}
    \caption{Scaled MFPT as a function of chirality ($\chi$) for CABPs in a 1D interval with reflecting left and absorbing right boundaries, computed using FEM. We plot $T/T_\mathrm{PBP}$ for particles starting at $x = 0$ under three initial-orientation conditions: (a) $\phi = 0$ (right-pointing), (b) $\phi = \pi$ (left-pointing), and (c) $\phi$ uniformly random, and for three swim speed: $Pe = 1$ (red circles), $Pe = 10$ (blue diamonds), and $Pe = 20$ (green squares).
 In each case, $T_{\mathrm{PBP}}$ is computed from Eq.~\eqref{eq:Tbar-passive_Ref_Abs_BC}. For all results, $\beta=0.1$. Note the differing $y$-axis limits in panels (a)--(c). The dashed lines in panel (c) indicate the high-chirality asymptotic solution given in Eq.~\eqref{eq:high-chi-midpoint-left_reflect}. }
    \label{fig:left_reflect_num}
\end{figure*}

Using the bulk and boundary-layer solutions, we construct a composite asymptotic solution for the MFPT of CABPs in the interval, given by
\begin{equation}
\label{eq:reflect-composite}
\begin{split}
      \Tbar_c = &~\Tbar_0(\xbar,\phi) + \frac{1}{\chi}\left( \Tbar_1(\xbar,\phi) +G_1\right) \\ 
      & ~+\frac{1}{\chi^{3/2}} \left( \Tbar_{3/2}(\xbar,\phi) + G_{3/2} \right) + O(1/\chi^2), 
\end{split}
\end{equation}
where $\Tbar_0(\xbar,\phi)$ is given in Eq.~\eqref{eq:Tbar-passive_Ref_Abs_BC},  $\Tbar_1(\xbar,\phi)$ is given in Eq.~\eqref{eq:T1-bulk-leftreflect},  and $\Tbar_{3/2}(\xbar,\phi)$ is given in Eq.~\eqref{eq:T32-reflect-sol}.  Here, $G_1$ and $ G_{3/2}$ are defined as follows:
\begin{equation}
    G_1 = -2Pe\;e^{-\lambda s} \sin(\lambda s +\phi),
\end{equation}
and 
\begin{equation}
\begin{split}
        G_{3/2}=&~\lambda  H(\lambda  z,\phi ) +Pe\lambda H(\lambda s, 0)  \\ 
        & ~+ Pe\lambda \Big[  H( s \sqrt{Pe}, 2\phi) - H(\lambda s, 2\phi) \Big]. 
\end{split}
\end{equation}

In Fig.~\ref{fig:1d-reflect-theory}(a), we compare $(T-T_{\mathrm{PBP}})/\tau_s$ 
obtained from the full FEM solution against the asymptotic prediction, given by 
$\left(\bar{T}_1 + G_1\right)/\chi + \left(\bar{T}_{3/2} + G_{3/2}\right)/\chi^{3/2}$ 
via Eq.~\eqref{eq:reflect-composite}. The asymptotic solution shows excellent agreement 
with the FEM result across the entire $x$-domain for all values of $\phi$ considered. 
As in the case with absorbing boundaries at both ends, we observe a non-monotonic dependence on $\xbar$. Similarly, as $\phi$ increases from negative to positive values, CABPs undergo a continuous transition from facilitated to hindered transport regimes (cf. Fig.~\ref{fig:1d-absorbing-theory}).

In Fig.~\ref{fig:1d-reflect-theory}(b), we show a contour plot of the MFPT for CABPs obtained from the three-term asymptotic solution in Eq.~\eqref{eq:reflect-composite} as a function of the initial position and orientation of the particles, for $\chi=50$ and $Pe=1$. In the high-chirality regime, the MFPT of CABPs is, to leading order, identical to that of PBPs [see first term in Eq.~\eqref{eq:reflect-composite}]. 
The presence of higher-order correction terms introduces a weak dependence of the MFPT on the particle orientation $\phi$. 

Since the left boundary is reflecting, the MFPT attains its maximum there;  evaluating the composite solution at $\bar{x} = -1$ yields
\begin{equation}
    \Tbar(-1,\phi) = 2 Pe + \frac{\lambda}{\chi^{3/2}}\left( -2 Pe +  \sin\phi+\cos\phi \right) + O\left( 1/\chi^2\right).
\end{equation}
Notably, the $O(1/\chi)$ terms in Eq.~\eqref{eq:reflect-composite} vanish at $\bar{x} = -1$, so the leading correction to the passive result appears at $O\left( 1/\chi^{3/2}\right)$. Furthermore, the MFPT of CABPs at the reflecting boundary is slighted reduced compared to that of PBPs, thanks to the coupling of swimming and chiral rotation assisting the particle steering away from the wall.

\subsubsection{The finite-chirality regime: reflecting-absorbing boundaries}

As in the absorbing-boundary problem, the MFPT for CABPs in the interval with a left reflecting boundary admits no closed-form solution in the finite-chirality regime. We therefore solve Eq.~\eqref{eq:T-non-dim-1D} subject to the boundary conditions in Eq.~\eqref{Eq:BC_reflect_Absorb} using FEM.

\begin{figure*}[!t]
    \centering
    \includegraphics[width=5.1in]{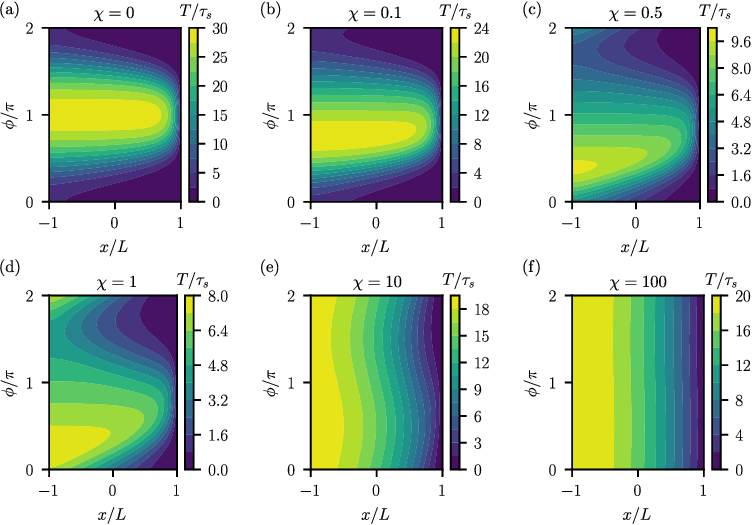}
    \caption{Contour plots showing the MFPT of CABPs in a 1D interval with left reflecting and right absorbing boundaries as a function of the initial position ($x/L$) and orientation ($\phi/\pi$) of the particles for varying chirality $\chi$. Panels show results for (a) $\chi = 0$, (b) $\chi = 0.1$, (c) $\chi = 0.5$, (d) $\chi = 1$, (e) $\chi = 10$, and (f) $\chi = 100$. In all cases, $Pe = 10$ and $\beta = 0.1$. }
    \label{fig:1d-contour}
\end{figure*}

The computed MFPT for CABPs is presented in Figs.~\ref{fig:left_reflect_num} and \ref{fig:1d-contour}. Fig.~\ref{fig:left_reflect_num} shows the normalized MFPT, $T/T_\mathrm{PBP}$, as a function of chirality, for particles starting at the midpoint of the interval ($\xbar = 0$) with varying initial orientations and different P\'eclet numbers. 
In Fig.~\ref{fig:left_reflect_num}(a), we show the MFPT for particles initially oriented to the right ($\phi = 0$). The MFPT profile closely resembles that of the symmetric interval with absorbing boundaries at both ends [see Fig.~\ref{fig:mfpt-1d-absorbing}(a)], increasing monotonically with $\chi$. Although the ratios in Figs.~\ref{fig:mfpt-1d-absorbing}(a) and \ref{fig:left_reflect_num}(a) suggest that $T/T_\mathrm{PBP}$ is lower for the interval with a left reflecting boundary, this does not imply that the absolute MFPT is smaller, since the passive MFPTs for the two domain configurations differ.

So far, the MFPT computed across all scenarios and both 1D domain configurations as a function of $\chi$ shows that the MFPT for CABPs is generally less than or equal to that of PBPs [see Fig.~\ref{fig:mfpt-1d-absorbing}(a)--(c) and Fig.~\ref{fig:left_reflect_num}(a)]. However, Fig.~\ref{fig:left_reflect_num}(b) presents a contrasting case: when particles start oriented toward the reflecting boundary (left-pointing), the MFPT for CABPs is greater than that of PBPs for all values of $Pe$ when $\chi$ is very small, as indicated by $T/T_\mathrm{PBP} > 1$. In this small-$\chi$ regime, the escape is dominated by the achiral active Brownian dynamics. ABPs initially oriented towards the left wall become trapped there and must rely on rotational Brownian diffusion to reorient away from the wall. As $\chi$ increases, the MFPT decreases for all values of $Pe$. For $Pe = 1$ (red circles), this decrease is monotonic, and the MFPT converges to that of PBPs as $\chi \to \infty$.  
For higher swim speeds, $Pe = 10$ and $Pe = 20$, the MFPT exhibits non-monotonic behavior: it initially decreases as $\chi$ increases, reaches a minimum, and then increases again with further increases in $\chi$, eventually converging to the MFPT for PBPs as $\chi \to \infty$. For these larger values of $Pe$, the MFPT for CABPs drops below that of PBPs over a range of $\chi$.

In Fig.~\ref{fig:left_reflect_num}(c), we show the results for particles with uniformly random initial orientations, which effectively average the behaviors observed in Figs.~\ref{fig:left_reflect_num}(a) and \ref{fig:left_reflect_num}(b). In this case, the MFPT exhibits a minimum at an intermediate value of $\chi$. Unlike the symmetric interval with uniformly random orientations [cf. Fig.~\ref{fig:mfpt-1d-absorbing}(c)], where CABPs consistently have MFPTs below the passive MFPT ($T/T_\mathrm{PBP} < 1$) for all $Pe$, in the left reflecting boundary configuration, the MFPT for CABPs exceeds that of PBPs at $Pe = 1$, but is lower for higher swim speeds ($Pe = 10$ and $Pe = 20$). For all swim speeds, the variation in $T/T_\mathrm{PBP}$ with increasing $\chi$ is more pronounced for the reflecting-left boundary configuration, highlighting the influence of boundary type on the MFPT as a function of chirality.

In the high-chirality regime, applying Eq.~\eqref{eq:reflect-composite} at the domain midpoint ($\bar{x} = 0$) yields the asymptotic expansion
\begin{equation}
        \frac{T}{T_\mathrm{PBP}}\Big \rvert_{\xbar=0} = 1 + \frac{1}{\chi}\frac{2\sin\phi}{3} - \frac{\lambda}{\chi^{3/2}} +O\left(1/\chi^2\right). 
    \end{equation}
We note that the order of the leading correction to the passive result depends on the initial orientation angle. When $\sin\phi$ = 0, as shown in Figs.~\ref{fig:left_reflect_num}(a)--(b), the first nonzero correction arises at $O\left(1/\chi^{3/2}\right)$; for $\sin\phi \neq 0$, it appears at $O(1/\chi)$. For CABPs with an initially uniform orientational distribution, the $O(1/\chi)$ term vanishes, yielding 
\begin{equation}
\label{eq:high-chi-midpoint-left_reflect}
            \frac{T_\mathrm{uni}}{T_\mathrm{PBP}}\Big \rvert_{\xbar=0} = 1  - \frac{\lambda}{\chi^{3/2}} +O\left(1/\chi^2\right). 
\end{equation}
This asymptotic result is shown as dashed lines in Fig.~\ref{fig:left_reflect_num}(c), exhibiting excellent agreement with the numerical results in the high-chirality regime.

It is instructive to compare $T_\mathrm{uni}$ for particles starting from the midpoint ($\xbar=0$) under two boundary conditions: both ends absorbing, and a reflecting left boundary with an absorbing right boundary. While $T_\mathrm{PBP}(\xbar=0)$ differs between the two cases, the two-term asymptotic solution for the scaled MFPT, $T_\mathrm{uni}(\xbar=0)/T_\mathrm{PBP}(\xbar=0)$  coincides for both boundary conditions [cf. Eqs.~\eqref{eq:high-chi-midpoint-left_reflect} and \eqref{eq:absorbing-large-chi-scaled}]. We note that this is not generally true for an arbitrary $\xbar$, which reflects precisely the interplay between asymmetric confinement, chirality and activity.

Next, we present contour plots  of the MFPT for CABPs in the 1D interval with a left reflecting boundary, as functions of the initial position and orientation of the particles in Fig.~\ref{fig:1d-contour}. The physical parameters are $Pe=10$ and $\beta=0.1$.
Each subplot corresponds to a different fixed value of chirality $\chi$. For $\chi=0$ [see Fig.~\ref{fig:1d-contour}(a)], the MFPT corresponds to that of standard ABPs.  In this case, the MFPT is maximized when particles are initially oriented toward the reflecting boundary and located at $x/L = -1$ (i.e., $\phi/\pi = 1$), and it decreases symmetrically as the initial orientation deviates from this direction. Generally, for a fixed initial orientation, the MFPT decreases as the starting position moves closer to the absorbing boundary at $x/L = 1$. This behavior is intuitive, as particles near the absorbing boundary are more likely to escape quickly. We further note that for $\phi = \pi$, ABPs initially swim toward the wall before reorienting and eventually exiting. If the P\'eclet number is large, the transit time from the bulk to the left wall is short, and the MFPT is therefore dominated by the reorientation process at the wall. Consequently, for large $Pe$ (e.g., $Pe = 10$) and $\phi=\pi$, $T/\tau_s$ exhibits a plateau before decreasing to zero, consistent with the absorbing boundary condition at the right exit  (see also Fig.~\ref{fig:ABP} in Appendix~\ref{app:figures}).

As $\chi$ increases from zero, the particles begin to spin in the counterclockwise direction (since $\chi > 0$). This rotational motion significantly affects their overall dynamics, as reflected in the MFPT required to escape the domain [see Fig.~\ref{fig:1d-contour}(b)--(e)]. Notably, for small $\chi$ [see Fig.~\ref{fig:1d-contour}(b); see also Fig.~\ref{fig:ESIphi} in Appendix~\ref{app:figures}], the previously symmetric contour  begins to distort, and a pronounced asymmetry emerges as $\chi$ increases further [see Fig.~\ref{fig:1d-contour}(c)--(d)]. Owing to the counterclockwise rotation, the location of the maximum MFPT at the left wall shifts to orientations with $\phi < \pi$.   This arises because particles starting with $\phi < \pi$ are advected by chirality toward the left wall, which increases their residence time before reaching the escape-favorable orientation window $|\phi| < \pi/2$. On the other hand, for $\phi > \pi$, rotational motion carries particles directly into this escape window, allowing for faster escape. Additionally, the maximum MFPT [see the color bars in Figs.~\ref{fig:1d-contour}(a)–(d)] decreases relative to the $\chi = 0$ (ABP) case, indicating that rotational dynamics reduce the worst-case escape time. Physically, the worst-case scenario for ABPs corresponds to orientations pointing toward the left wall, where particles become effectively trapped by persistent propulsion. Chirality introduces rotational dynamics that break this trapping by reorienting particles, thereby enhancing escape and reducing the MFPT.

As mentioned earlier, when CABPs rotate rapidly, their directed motion becomes increasingly disrupted, and their behavior approaches that of PBPs, as $\chi \to \infty$. This transition is evident in the contour plots in Fig.~\ref{fig:1d-contour}(e) ($\chi = 10$) and \ref{fig:1d-contour}(f) ($\chi = 100$). For $\chi = 10$, the influence of the initial orientation on the MFPT is already significantly reduced compared to lower values of $\chi$. In this case, the escape dynamics is largely dominated by chirality, which is characterized in Section~\ref{subsubsec:reflect-absorb-theory}. For $\chi = 100$, effect of persistent propulsion becomes negligible.  In this regime, the MFPT is determined primarily by the particle's initial position, rather than its orientation or swim speed, which is characteristic of PBPs and consistent with the results for the symmetric domain shown in Fig.~\ref{fig:absorbing-contour}(d), and our asymptotic analysis.

Lastly, we note from Figs.~\ref{fig:mfpt-1d-absorbing} and \ref{fig:left_reflect_num} that the MFPT for CABPs decreases as the particle swim speed increases, i.e., as $Pe$ increases. This trend is expected, as higher activity enhances the ability of particles to escape from the domain regardless of boundary type, consistent with previous findings for ABPs~\cite{iyaniwura2024asymptotic}. Although this is not immediately evident from the computed ratio $T/T_\mathrm{PBP}$ in all cases, the MFPT for CABPs is higher in the interval with a left reflecting boundary than in the symmetric domain with absorbing boundaries at both ends. This is also expected, since particles in the symmetric domain can exit from either boundary, whereas those in the domain with a left-reflecting boundary can only exit through the right boundary. This difference is apparent when comparing the  results in Figs.~\ref{fig:1d-absorbing-theory}(b) and~\ref{fig:absorbing-contour} with Figs.~\ref{fig:1d-reflect-theory}(b) and~\ref{fig:1d-contour}.

\begin{figure}[!t]
    \centering
    \includegraphics[width=3.0in]{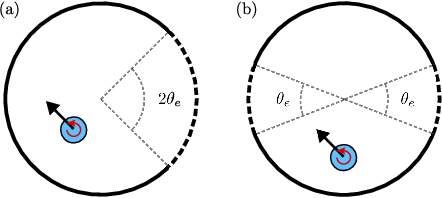}
    \caption{
Schematic illustration of a CABP confined within a circular domain with absorbing arcs (dashed lines) through which the particle can escape. The remaining portions of the boundary (solid lines) are reflecting.
(a) A CABP in a disk with a single absorbing arc spanning the angular interval $[-\theta_e, \theta_e]$.
(b) A CABP in a disk with two symmetric absorbing arcs located at $\theta \in (-\theta_e/2, \theta_e/2)$ and $\theta \in (\pi - \theta_e/2, \pi + \theta_e/2)$.
The blue circle represents the CABP, the black arrow indicates its instantaneous orientation (self-propulsion direction), and the red curved arrow denotes chirality.
}
\label{fig:disk_sketch}
\end{figure}

\section{CABPs in a disk}\label{sec:MFPT_Disk}

We extend our analysis to the escape dynamics of CABPs in a two-dimensional disk of radius $R$. The MFPT in this geometry remains governed by Eq.~\eqref{eq:MFPT-EQ}. We use $R$ as the length scale of the domain so that the swim timescale is $\tau_s = R/U_s$. Non-dimensionalizing Eq.~\eqref{eq:MFPT-EQ} using this timescale yields the dimensionless 
parameters:
\begin{equation}\label{eq:dimless_par_2D}
    Pe = \frac{U_s R}{D_x}, \qquad \chi = \frac{\Omega R}{U_s}, \qquad 
    \beta = \frac{R}{U_s \tau_R}.
\end{equation}
In analyzing the MFPT in this geometry, we consider two configurations: (i) a disk with a single absorbing arc, and (ii) a disk with two symmetric absorbing arcs (Fig.~\ref{fig:disk_sketch}). For each configuration, the MFPT is computed using FEM  as a 
function of the P\'{e}clet number ($Pe$) and chirality ($\chi$).

\subsection{CABPs in a disk with one absorbing arc}\label{sec:MFPT_Disk_1Arc}

\begin{figure}[!t]
    \centering
    \includegraphics[width=3in]{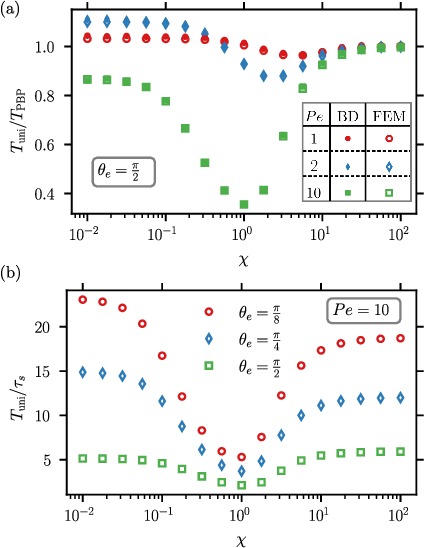}
    \caption{MFPT for CABPs in a 2D disk with a single absorbing arc as a function of the angular velocity ($\chi$). (a) $\theta_e=\pi/2$ and $Pe=1,2,10$, and (b) $Pe=10$ and $\theta_e = \pi/8, \pi/4, \pi/2$. For all scenarios, CABPs start at the center of disk with uniformly random initial orientation $(\phi)$ and $\beta=0.1$. $T_\mathrm{PBP}$ is the passive MFPT and $\tau_s$ is the swimming time scale of the particles. In (a), MFPT computed from BD simulations are shown using filled markers. }
    \label{fig:mfpt-disk-1exit}
\end{figure}

Consider a CABP confined to a disk of radius $R$ with an absorbing arc between the angles $-\theta_e$ and $\theta_e$, through which the particle can escape [see Fig.~\ref{fig:disk_sketch}(a)].  The remainder of the disk boundary is reflecting. The MFPT for the particle to exit the disk satisfies Eq.~\eqref{eq:MFPT-EQ}. In non-dimensional form, we solve 
\begin{equation}
\label{eq:MFPT-2D-nondimensional}
    \bq\cdot\bar{\nabla} \Tbar + \frac{1}{Pe}\bar{\nabla}^2 \Tbar + \chi \frac{\partial \Tbar}{\partial \phi } + \beta \frac{\partial ^2 \Tbar}{\partial \phi^2}=-1, 
\end{equation}
subject  to the Dirichlet boundary condition
\begin{equation}
\label{eq:2D-absorb-bc}
  \Tbar(\bar{r}=1,\theta,\phi) = 0 \quad \text{for} \quad \theta \in \Theta_e ,
\end{equation}
on the absorbing arc, and the Neumann boundary condition 
\begin{equation}
\label{eq:2D-reflect-BC}
    \frac{\partial \Tbar}{\partial \bar{r}}(\bar{r}=1,\theta,\bq) = 0 \quad \text{for } \theta \in [0,2\pi] \setminus \Theta_e,
\end{equation}
on the reflecting portion of the boundary. Here, $\Theta_e \equiv (-\theta_e,\theta_e)$ denotes the absorbing arc, $\bar{r}=r/R$, and $\bar{\nabla} = R \nabla$ is the non-dimensional gradient operator. As in the 1D cases, $\bq = \cos\phi\;\be_x + \sin\phi \be_y$.

The MFPT for a CABP starting at the center of the disk with a uniformly random initial orientation is shown in Fig.~\ref{fig:mfpt-disk-1exit}, plotted as a function of chirality $\chi$ for different values of $Pe$. These results extend the analysis of Section~\ref{sec:MFPT_1D}, which considered a CABP in a 1D interval with a 
reflecting boundary on the left and an absorbing boundary on the right, to the 2D setting. In particular, the configuration shown in Fig.~\ref{fig:mfpt-disk-1exit}(a) for $\theta_e = \pi/2$ is analogous to that in 
Fig.~\ref{fig:left_reflect_num}(c) for the 1D case.

As in the 1D case, there exists an intermediate value of $\chi$ that minimizes the MFPT. This minimum is sharp at large P\'eclet numbers, as shown for $Pe=10$ in Fig.~\ref{fig:mfpt-disk-1exit}(a). In the large-$Pe$ regime, the minimum is set by the balance $R/U_s \approx 1/\Omega$, which again gives $\chi \approx 1$. As $Pe$ decreases, translational Brownian motion becomes increasingly important, and the minimum becomes both shallower and broader. This behavior reflects a generic trade-off: at low $Pe$, diffusive transport reduces the sensitivity of the MFPT to the chiral dynamics set by $\chi$. Put differently, the reduction in MFPT due to optimal chirality is most pronounced for highly active particles.

For ABPs or CABPs with small $\chi$, the escape dynamics  are effectively governed by standard ABP behavior. In contrast to the 1D case---where the MFPT decreases monotonically with increasing $Pe$---the 2D results exhibit a  more intricate dependence on $Pe$. Specifically, for small or vanishing $\chi$, the MFPT is
larger for $Pe = 2$ than for $Pe = 1$. For ABPs, this non-monotonic dependence of the MFPT on $Pe$ has been discussed in Iyaniwura \& Peng~\cite{iyaniwura2025mean}; the MFPT distribution as a function of $Pe$ and rotational Brownian motion has also been studied theoretically in a circular region with the boundary being entirely absorbing~\cite{di2023active}.

To examine the dependence of the MFPT on the size of the absorbing arc, in Fig.~\ref{fig:mfpt-disk-1exit}(b) we show the MFPT for several values of the absorbing arc half-angle $\theta_e$ ($\theta_e = \pi/8, \pi/4, \pi/2$) at a fixed value of $Pe=10$. Increasing $\theta_e$ corresponds to enlarging the absorbing arc, thereby increasing the likelihood that a  particle encounters the escape window along the boundary; as a result, the MFPT decreases monotonically with increasing $\theta_e$. As discussed earlier, for large $Pe$, the optimality arises from the balance $R/U_s \approx 1/\Omega$. Consequently, the chirality that minimizes the MFPT is largely independent of the size of the absorbing arc. Therefore, the drop in the MFPT becomes more pronounced with a smaller window size.

Similar to the 1D case [Fig.~\ref{fig:left_reflect_num}(c)] is that the MFPT of CABPs, and of ABPs in the small chirality limit, can be shorter or longer than that of PBPs as shown in Fig.~\ref{fig:mfpt-disk-1exit}. This again reflects the competition previously described:  particles with unfavorable initial orientations are more likely to encounter the reflecting boundary first and may even become trapped, whereas those oriented toward the exit can escape more rapidly due to active swimming. This competition is further  modulated by the swimming speed $Pe$ and the size of the opening $\theta_e$.

\subsection{CABPs in a disk with two symmetric absorbing arcs}\label{sec:MFPT_Disk_2SymmArc}

\begin{figure}[!t]
    \centering
    \includegraphics[width=3in]{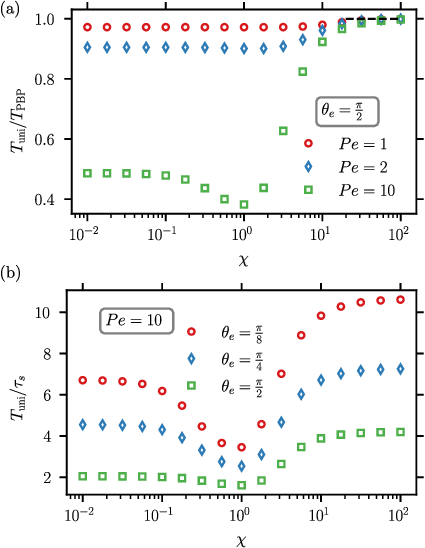}
    \caption{MFPT for CABPs in a 2D disk with two symmetric absorbing arcs as a function of the angular velocity ($\chi$). (a) $\theta_e=\pi/2$ and $Pe=1,2,10$, and (b) $Pe=10$ and $\theta_e = \pi/8, \pi/4, \pi/2$. For all scenarios, CABPs start at the center of disk with uniformly random initial orientation $(\phi)$ and $\beta=0.1$. $T_\mathrm{PBP}$ is the passive MFPT and $\tau_s$ is the swimming time scale of the particles.}
    \label{fig:mfpt-disk-2exits}
\end{figure}

Next, we consider a disk with two symmetric absorbing arcs located at angles $\theta \in  (-\theta_e/2,\theta_e/2)$ and $ \theta \in (\pi-\theta_e/2,\pi+\theta_e/2)$,
through which a CABP can escape. The remainder of the boundary is reflecting [see Fig.~\ref{fig:disk_sketch}(b)]. Let $\Theta_e \equiv (-\theta_e/2,\theta_e/2) \cup  (\pi-\theta_e/2,\pi+\theta_e/2) $, the MFPT for a CABP to exit the disk satisfies Eq.~\eqref{eq:MFPT-2D-nondimensional}, with the boundary conditions given by Eqs.~\eqref{eq:2D-absorb-bc} and \eqref{eq:2D-reflect-BC}. We note that the only difference between this case and Sec.~\ref{sec:MFPT_Disk_1Arc} in the PDE formulation lies in the definition of $\Theta_e$. We choose the absorbing arc $\Theta_e$ such that the total length equals that of the single absorbing arc in Fig.~\ref{fig:disk_sketch}(a). This geometry is analogous to the 1D interval with absorbing boundaries at both ends, analyzed in Section~\ref{subsec:absorbing-1d}.

As shown in Fig.~\ref{fig:mfpt-disk-2exits}(a), for $Pe=10$, the MFPT exhibits a non-monotonic dependence on $\chi$. In contrast to the one-arc case, this non-monotonicity is largely absent at lower $Pe$, as illustrated for $Pe = 1$ and $Pe = 2$ in Fig.~\ref{fig:mfpt-disk-2exits}(a). With two exits, particles can escape from either side, which reduces geometric constraints on the dynamics, similar to the 1D cases. As a result, only highly active particles—those that tend to become trapped along the reflecting portions of the boundary—benefit significantly from an optimal chirality.

Lastly, Fig.~\ref{fig:mfpt-disk-2exits}(b) shows the MFPT for a CABP escaping the disk for $Pe=10$ and $\theta_e = \pi/8, \pi/4, \pi/2$. As in the single-arc case, the MFPT decreases monotonically with increasing arc length, and the optimal $\chi$ that minimizes the MFPT is largely independent of the absorbing arc length. However, the dependence on the chirality $\chi$ differs qualitatively from the single-arc geometry. Specifically, for 
a disk with one absorbing arc, the MFPT is smaller in the limit $\chi \to \infty$ than in the limit $\chi \to 0$ for all values of $\theta_e$ considered. In contrast, for two symmetric absorbing arcs, the MFPT is lower as $\chi \to 0$ than as $\chi \to \infty$.

Since the dynamics of CABPs approach those of ABPs as $\chi \to 0$ and those of PBPs as $\chi \to \infty$, these results imply that ABPs escape more slowly than PBPs in the single-arc geometry, but more rapidly in the two-arc geometry. In the single-arc case, ABPs tend to become trapped near the reflecting boundary, which increases their residence time in the domain before reaching the exit. In contrast, with two exits, they can reach the boundaries more efficiently; sliding along the reflecting segments on either side eventually leads to an exit, enhancing escape efficiency~\cite{iyaniwura2026splitting}.  Note that this is similar to the contrast between 1D cases with absorbing/absorbing and reflecting/absorbing boundaries.

\section{Discussion}

In this work, we studied the MFPT of CABPs in confined geometries, considering two canonical configurations in both 1D and 2D settings. In 1D, we examined a particle in an interval with absorbing boundaries at both ends, and an interval with a reflecting boundary and an absorbing boundary. The 2D analysis considered analogous configurations on a disk: a disk with two symmetric absorbing arcs, corresponding to the 
fully absorbing 1D interval, and a disk with a single absorbing arc, corresponding to the mixed boundary 1D interval. By combining FEM solutions of the full problem with matched asymptotic expansions in the high-chirality regime, we obtained a comprehensive picture of how chirality, swim speed, and Brownian motion collectively govern the escape dynamics of CABPs.

A central finding of this work is the existence of an optimal chirality that minimizes the MFPT, a feature that is especially pronounced for CABPs at large P\'eclet numbers. Primarily for CABPs with unfavorable initial orientations, the optimality arises when the time of reorientation towards the exit, thanks to chirality, matches the time to transit the confinement domain through active swimming. It is generally independent of geometric dimensions (1D or 2D) and escape configurations (single or double symmetric exits).
This universality highlights chirality as a robust control parameter that can be 
tuned to minimize escape times across a broad class of confined geometries, with practical 
implications for the design of synthetic microswimmer systems in which the rotation rate 
can be controlled externally, for instance through magnetic fields or geometric 
confinement~\cite{ghosh2009controlled, zhang2009artificial, dreyfus2005microscopic}. This perspective draws a natural connection to studies of microswimmers in 
corrugated or periodically structured environments, where boundary curvature and shape 
have been shown to modulate swimmer trajectories and 
transport~\cite{modica2022porous,modica2023boundary,malgaretti2023splitting, baouche2026spatiotemporal, iyaniwura2026splitting}. 
The present results suggest that chirality introduces a control parameter in addition to boundary curvature. The interplay between chirality and boundary geometry can therefore give rise to non-trivial transport behavior~\cite{chan2024chiral}.

In the high-chirality regime ($\chi \gg 1$), the MFPT approaches that of PBPs to leading 
order. In 1D using asymptotic expansions, we have systematically resolved the structure of CABP escape time compared to its PBP limit, uncovering a delicate interaction among chirality, swim-induced escape, and Brownian diffusion. Meanwhile, this recovery of passive-like behavior reflects the fact that 
strong chirality suppresses persistent swimming and reduces net directed transport. Thus, while 
intermediate chirality enhances escape efficiency, excessive chirality is detrimental, 
underscoring the importance of tuning chirality to its optimal value.

The initial orientation $\phi$ of the particles plays a qualitatively distinct role in shaping the MFPT depending on the boundary configuration. In the fully absorbing case, the MFPT transitions smoothly between facilitated escape, when the 
particle is initially directed toward either exit and self-propulsion immediately drives it toward an absorbing 
boundary, and hindered escape, when it starts perpendicular to the boundaries and must first reorient before making 
productive progress toward an exit, a process governed by rotational diffusion and therefore costly in time. In 
the mixed boundary case, the maximum MFPT occurs when the particle starts pointing toward the reflecting boundary, 
since self-propulsion initially drives it away from the only available exit, effectively increasing the distance 
the particle must cover before escaping; as chirality increases, the orientation associated with this maximum 
rotates continuously in the direction of the chiral angular drift, reflecting the systematic bias that chirality 
introduces into the particle's trajectory. In both cases, the sensitivity of the MFPT to $\phi$ diminishes 
 with increasing chirality and vanishes as $\chi \to \infty$: rapid chiral rotation causes the 
self-propulsion direction to precess rapidly through all angles, averaging out any directional bias over timescales 
short compared to the escape time, erasing the memory of the initial orientation, and recovering the isotropic MFPT 
of a passive Brownian particle independent of $\phi$.

In the 2D setting, the MFPT was analyzed numerically across both absorbing arc 
configurations. A key finding is that the configuration of absorbing arcs can qualitatively 
alter the escape dynamics beyond a simple rescaling, even when the total absorbing arc 
length is held fixed between the two configurations. In particular, unlike the 1D case where the MFPT varies 
monotonically with swim speed across all chirality values considered, the 2D single-arc 
configuration exhibits a non-monotonic dependence of the MFPT on swim speed at low chirality.  No such behavior was observed in the two-arc configuration, where the MFPT varies monotonically with swim speed across the full range of chirality values considered. These results demonstrate that the number and placement of absorbing arcs in 2D or 3D can qualitatively alter the relationship between swim speed and escape time, even when the total absorbing arc length is held fixed. Unlike in 1D, particles in 2D or 3D are more likely to remain trapped for longer at a reflecting boundary. In a 1D interval, even a small deviation from the normal orientation to the wall allows the particle to slide along the boundary and eventually escape. In higher dimensions, however, particles must undergo a more substantial reorientation before they can move away from the boundary and reach the exit. As an example, this can be seen by comparing Fig.~\ref{fig:mfpt-1d-absorbing}(c) and Fig.~\ref{fig:mfpt-disk-2exits}(b), where the non-monotonicity is more pronounced in the 2D case.

Furthermore, in both configurations, the optimal chirality that minimizes the MFPT is found to be independent of the absorbing arc length, suggesting that this optimality is a robust geometric feature of the escape dynamics rather than a boundary-size effect.
 However, the results for these two configurations differ qualitatively in the limiting behavior 
of the MFPT: for a single absorbing arc, the MFPT is smaller as $\chi \to \infty$ than as 
$\chi \to 0$, whereas for two symmetric absorbing arcs the opposite holds, with the MFPT 
being lower as $\chi \to 0$ than as $\chi \to \infty$. This qualitative reversal underscores the role of domain geometry in determining whether high or low chirality is more detrimental to escape.

Despite the in-depth analysis carried out in this work, several avenues remain open for 
future investigation. The present model assumes a uniform translational diffusion 
coefficient and a fixed swim speed; relaxing these assumptions to account for spatially 
varying diffusivity or chirality would broaden the applicability of the 
framework to experimentally relevant systems. A natural extension is also the case of 
multiple absorbing windows with arbitrary placement and size, beyond the symmetric two-arc 
configuration considered here, particularly in the context of modeling cell membrane 
transport or drug delivery in complex environments.

Another natural extension of this work is the narrow escape problem for a CABP confined 
to a 2D disk with a fully reflecting boundary, where the absorbing targets 
consist of one or more small traps of size $\epsilon \ll 1$ located in the interior of 
the disk~\cite{schuss2007narrow, cheviakov2010asymptotic, iyaniwura2021optimization}. A key question is whether closed-form  asymptotic solutions can be derived in this setting. Extending this analysis to CABPs would reveal how chirality and swim speed 
modify the asymptotic behavior of the MFPT as $\epsilon \to 0$, and whether the optimal 
chirality identified in the present work persists in this interior trap setting. This 
extension is relevant to applications such as the targeting of drug-carrying microswimmers 
to specific receptor sites, and the arrival of reactive molecules at localized catalytic 
sites within a confined cellular environment.

Overall, the results presented in this work establish a quantitative framework for 
understanding how chirality modulates escape dynamics in confined active matter systems. 
A key outcome is the identification of an optimal chirality that minimizes the MFPT, 
positioning chirality as a robust control parameter for modulating the motion of active 
particles. This shows that chirality  could be leveraged in 
synthetic microswimmer systems to achieve targeted transport, with potential applications 
in drug delivery, microfluidics, and biomedical engineering, and provides a foundation 
for future studies of escape dynamics in more complex environments relevant to biological 
and synthetic microswimmers.

\section*{Author contributions}
S.A.I. : Conceptualization, Formal analysis,  Writing – original draft, Writing – review \& editing;
M.Q. :  Conceptualization, Formal analysis, Writing – review \& editing;
Z.P. : Conceptualization, Formal analysis,  Writing – original draft, Writing – review \& editing.

\section*{Conflicts of interest}
There are no conflicts to declare.

\section*{Data availability}

No new data were generated in support of this study. The simulation scripts are openly available at the following DOI: \url{https://doi.org/10.5683/SP3/MKXKDU}

\begin{acknowledgments}
This work was enabled in part by the Digital Research Alliance of Canada. Z.P. was supported by the Natural Sciences and Engineering Research Council of
Canada (NSERC), Grant No. RGPIN-2025-05310. Z.P. acknowledges support from the Erskine Fellowship at the University of Canterbury, during which part of this work was completed.
\end{acknowledgments}

\appendix
\section{Appendix}
\subsection{1D derivation with a reflecting boundary\label{app:appendix}}
To derive a boundary condition for $\Tbar_{3/2}$, we need to first obtain the bulk solution at $O(1/\chi^2)$. From the regular expansion given in Eq.~\eqref{eq:bulk-expand-revised}, we obtain 
\begin{equation}
\frac{1}{Pe}\frac{\partial^2\Tbar_1}{\partial \xbar^2}  + \beta \frac{\partial^2 \Tbar_1}{\partial \phi^2} + \cos\phi \frac{\partial \Tbar_1}{\partial \xbar} + \frac{\partial \Tbar_2}{\partial \phi}=0.
\end{equation}
Inserting Eq.~\eqref{eq:T1-bulk-leftreflect} into the above equation and integrating yields
\begin{equation}
   \Tbar_2(\xbar,\phi)= b(\xbar)-\beta  \, Pe\, \cos \phi  \left(1+\xbar\right)+\frac{1}{2} Pe\, \cos ^2\phi, 
\end{equation}
where $b(\xbar)$ remains to be determined.

For the boundary layer near the reflecting wall on the left, the matching condition is then given by 
\begin{equation}
\label{eq:appendix-matching}
\begin{split}
        \Tbar =~& 2 Pe   -\frac{1}{\chi} \frac{Pe\, z^2 }{2} + \frac{1}{\chi^{3/2}} \left(Pe\, z\, \sin\phi + \Tbar_{3/2}(-1) \right)\\
        & + \frac{1}{\chi^2}\left( \frac{Pe\,\cos^2\phi}{2} + b(-1)+z\; \Tbar_{3/2}^\prime(-1) \right) + O \left( \frac{1}{\chi^{5/2}}\right), 
\end{split}
\end{equation}
where we have used the fact that $a(-1)=0$. Recall the boundary-layer expansion 
\begin{equation}
    \tilde{T} = 2Pe + \frac{1}{\chi} \tilde{T}_1(z,\phi) + \frac{1}{\chi^{3/2}} \tilde{T}_{3/2}(z,\phi) + \frac{1}{\chi^2}\tilde{T}_2(z,\phi) + \cdots, 
\end{equation}
where $\tilde{T}_1$ and $\tilde{T}_{3/2}$ are given by Eqs.~\eqref{eq:T01-lbl-reflect} and \eqref{eq:T32-reflect-lbl}, respectively.

At $O(1/\chi^2)$, we obtain 
\begin{equation}
    \frac{1}{Pe}\frac{\partial^2 \Tbar_2}{\partial z^2 } + \frac{\partial \Tbar_2}{\partial \phi }+ \cos\phi \frac{\partial \Tbar_{3/2}}{\partial z} + \beta \frac{\partial^2 \Tbar_1}{\partial \phi^2}=0,
\end{equation}
whose solution is given by 
\begin{equation}
\begin{split}
    \tilde{T}_2(z, \phi) &= b(-1) + \frac{1}{2} Pe \cos^2(\phi) + \frac{Pe \lambda}{2} z + \frac{Pe}{2} e^{-\lambda z} \cos(\lambda z) \\
    &\quad - \frac{1}{2} Pe \; e^{-\lambda z} \cos(\lambda z + 2\phi) \\ 
    &\quad+ \frac{Pe}{2\sqrt{2}} e^{-z\sqrt{Pe}} \cos( z\sqrt{Pe} + 2\phi).
\end{split}
\end{equation}
Comparing this solution with the matching condition given in Eq.~\eqref{eq:appendix-matching}, we conclude that $\Tbar_{3/2}^\prime(-1) = Pe \lambda/2$. We note that, at this stage,  $b(-1)$ remains undetermined. Integrating the BVP, $\Tbar_{3/2}^{\prime\prime}=0$, together with the boundary conditions $\Tbar_{3/2}(1)= -Pe\lambda $ and $\Tbar_{3/2}^\prime(-1)= Pe \lambda/2$, we obtain 
\begin{equation}
    \Tbar_{3/2}(\xbar) = \frac{Pe\lambda }{2}\left( \xbar -3\right).
\end{equation}

\subsection{Additional figures\label{app:figures}}

\begin{figure}[!t]
    \centering
    \includegraphics[width=3.3in]{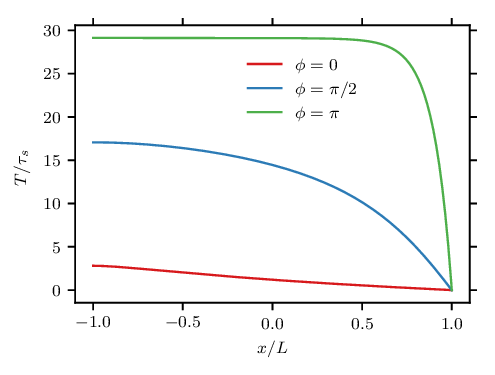}
    \caption{ MFPT of ABPs ($\chi=0$) in a 1D interval with reflecting left and absorbing right boundaries as a function of the initial position ($x/L$). Other parameters are $Pe=10$ and $\beta=0.1$. The corresponding contour plot for the MFPT as a function of $x/L$ and $\phi$ is given in Fig.~\ref{fig:1d-contour}(a).}
    \label{fig:ABP}
\end{figure}

\begin{figure}[!t]
    \centering
    \includegraphics[width=3.3in]{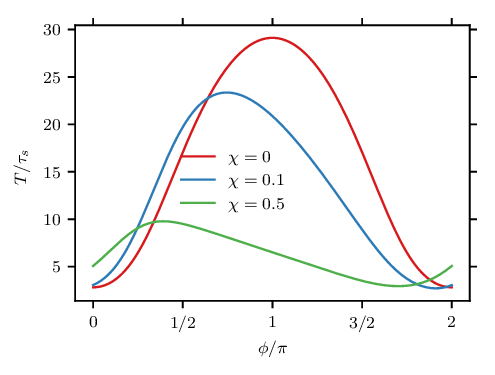}
    \caption{ MFPT of CABPs in a 1D interval with reflecting left and absorbing right boundaries as a function of the initial orientation. For all results, the starting particle position $x/L=-1$. That is, the particles start from the left reflecting wall. Other parameters are $Pe=10$ and $\beta=0.1$. See Figs.~\ref{fig:1d-contour}(a)--(c) for the corresponding contour plots for the MFPT as a function of $x/L$ and $\phi$. }
    \label{fig:ESIphi}
\end{figure}

In this section, we provide two additional figures (Figs.~\ref{fig:ABP} and \ref{fig:ESIphi}) illustrating the MFPTs of ABPs and CABPs in a one-dimensional interval with a reflecting boundary on the left and an absorbing boundary on the right. 

In Fig.~\ref{fig:ABP}, we plot the scaled MFPT, $T/\tau_s$, as a function of the position for different values of the orientation angle $\phi$. For $\phi=0$, the particle initially points towards the exit. As the starting position shifts from left to right, the MFPT decreases monotonically. For $\phi=\pi$, the particle initially points towards the reflecting wall. In this case, the MFPT exhibits a plateau before it eventually decreases to zero as $x/L$ approaches the exit. 

In Fig.~\ref{fig:ESIphi}, we plot the scaled MFPT, $T/\tau_s$, as a function of the initial orientation angle $\phi$ for different values of chirality. The initial particle position $x/L=-1$, i.e., the particle starts from the reflecting left wall. For ABPs ($\chi=0$), the maximum is attained  at $\phi=\pi$, corresponding to a particle initially oriented toward the reflecting left wall. In this configuration, the particle is initially stuck at the wall and must rely on rotational Brownian motion to reorient away from it. As chirality increases, the location of the maximum shifts towards smaller $\phi$.

\subsection{Brownian dynamics \label{app:BD}}

In a 1D spatial interval, the discretized Langevin equations are given by 
\begin{subequations}
\label{eq:BD-1d}
    \begin{equation}
        x_{n+1} =  x_n  + U_s\cos\left(\phi_n\right)\Delta t + \Delta x^B,
    \end{equation}
    \begin{equation}
        \phi_{n+1} = \phi_n + \Omega\; \Delta t+  \Delta \phi^B,
    \end{equation}
\end{subequations}
where $\Delta t$ is the time step.  The Brownian displacements  $\Delta x^B$,  $\Delta y^B$, and $\Delta \phi^B$ are sampled from independent white noise processes. The translational Brownian displacement has a variance of $2D_x\Delta t$, and the rotary Brownian displacement has a variance of $2\Delta t D_R$. 

For all simulations, a sufficiently small time step is used to resolve all the physical timescales in the system.  To ensure good statistics, all simulations are performed with $200,000$ particles. For each particle, the simulation is terminated upon reaching the absorbing boundary, and the corresponding first-passage time is recorded. The mean first passage time can be easily computed by an appropriate ensemble average. 

For CABPs in a 2D disk, in addition to Eq.~\eqref{eq:BD-1d}, the equation of motion for the $y$-coordinate reads
    \begin{equation}
        y_{n+1} =  y_n  + U_s\sin\left(\phi_n\right)\Delta t + \Delta y^B.
    \end{equation}

\balance 
\bibliography{rsc} 

\end{document}